\pdfoutput=1
\documentclass[a4paper, 12pt]{article}
\usepackage[utf8]{inputenc}
\usepackage[top=1in, bottom=1in, left=1in, right=1in]{geometry}
\usepackage[english]{babel}
\usepackage{graphicx}
\usepackage{amsthm}
\usepackage{subfig}
\usepackage{amsmath,amssymb}
\usepackage{framed}
\usepackage{minibox}
\usepackage{float}
\usepackage{braket}
\usepackage{slashed}
\usepackage{cases}
\usepackage{physics}
\usepackage{dsfont}
\usepackage{array}
\usepackage{hhline}
\usepackage{verbatim}
\usepackage{empheq}
\usepackage{mathrsfs}
\usepackage{mathtools}
\usepackage{xcolor}
\usepackage{layout}
\usepackage[colorlinks,allcolors=blue]{hyperref}

\newcolumntype{L}{>{$}l<{$}}

\newcommand{\p}{\partial}
\newcommand{\non}{\nonumber\\}
\newcommand{\beq}{\begin{eqnarray}}
\newcommand{\eeq}{\end{eqnarray}}
\newcommand{\TeV}{{\rm TeV}}
\newcommand{\GeV}{{\rm GeV}}
\renewcommand{\d}{{\mathrm{d}}}
\renewcommand{\i}{\mathrm{i}}

\begin{document}

\title{
\begin{flushright}\ \vskip -1.5cm {\small {IFUP-TH-2023}}\end{flushright}
\vskip 20pt
\bf{ Aspects of the Electroweak Skyrmion   }
}

\vskip 60pt  

\author{  
Stefano Bolognesi$^{(1)}$, Sven Bjarke Gudnason$^{(2)}$, 
and  Giacomo Santoni$^{(3)}$\\[13pt]
{\em \footnotesize
$^{(1)}$ Department of Physics ``E. Fermi", University of Pisa and INFN, Sezione di Pisa}\\[-5pt]
{\em \footnotesize
Largo Pontecorvo, 3, Ed. C, 56127 Pisa, Italy}\\[2pt] 
{\em \footnotesize  $^{(2)}$Institute of Contemporary Mathematics, School of
  Mathematics and Statistics,}\\[-5pt]
{\em \footnotesize Henan University, Kaifeng, Henan 475004,
  P.~R.~China }\\[2pt]
{\em \footnotesize  $^{(3)}$ University of Rome La Sapienza and INFN, Sezione di Roma} \\[-5pt]
{\em\footnotesize p.le Aldo Moro 2, Ed. Fermi, 00185 Rome, Italy }\\[4pt]
{\footnotesize  stefano.bolognesi(at)unipi.it, gudnason(at)henu.edu.cn, giacomo.santoni(at)uniroma1.it}
}
\date{}

\maketitle

\begin{abstract}
We consider certain aspects of the electroweak Skyrmion (EWS). 
We discuss the case of EWS with dynamical Higgs and find numerical
solutions for various values of the cutoff scale. Our results are
qualitatively similar to the ones present in the literature, but we
find a considerable lower mass than previous studies.
We discuss the quantization of the light degrees of freedom and prove that the EWS is a boson. 
We consider the interaction between fermions and the EWS and the
transfer of fermionic charge onto the soliton. 
We consider the large distance structure of the soliton and the
interaction between two well separated EWSs.  
We find that the classical EWS has a magnetic dipole moment.
We discuss the lifetime of the metastable soliton.
Finally, we discuss some phenomenological and cosmological
consequences of our results. 
\end{abstract}

\newpage
\tableofcontents

\section{Introduction}
Solitons are local minima of the static energy functional of a field theory.
Each individual soliton of a given class is labeled by a set of coordinates that describe its collective degrees of freedom e.g.~its position and orientation in space. When they are quantized, they generate a Hilbert space that can be decomposed into irreducible representations of the Galileo group and of the internal symmetry group of the theory. Hence, in the low-energy limit, soliton states behave as one-particle states. Furthermore, when fermions are coupled to the constituent fields of a soliton, the soliton may acquire a
fermion number, which may be non-integer and whose fractional part can be computed perturbatively. Several solitonic field configurations have been studied in the literature. Some of them are protected by a topological conservation law, that forbids their decay into the perturbative vacuum, both at the classical and at the quantum level. One of the field configurations
that enjoy topological protection is the \textit{Skyrmion}, a soliton of the chiral Lagrangian theory that was found \cite{Skyrme:1961vq} in an attempt to formulate a unified theory of mesons and baryons before the quark
model was available. The Skyrme model was then revived starting in Ref.~\cite{Adkins:1983ya}, with a work aimed to study their quantum properties. It was found that the predictions of the Skyrme model agree in general with about a 30\% or better accuracy with the experimental data on baryons.  Not all solitons have topological quantum numbers to protect them against decay.  Some of them are just metastable states.

The electroweak Skyrmion (EWS) is a Skyrmion in the Weinberg-Salam electroweak (EW) theory that appeared in the literature for the first time in Ref.~\cite{DHoker:1984wbw}.
The authors assumed the presence of a new non-renormalizable term, that must be added to the electroweak Lagrangian in order to stabilize the topologically nontrivial solutions. Two years later, two articles, \cite{Eilam:1985tg} and \cite{Ambjorn:1984bb}, studied numerically this possibility.
The existence of multi-Skyrmion solutions was studied in Ref.~\cite{Brihaye:1989ej}, and the classical process of production and destruction of EWSs was studied numerically in Ref.~\cite{Farhi:1995aq}. All these works made the simplifying assumptions of a non-dynamical Higgs scalar and of a decoupled hypercharge field. The topic was revived recently in Ref.~\cite{Criado:2020zwu}, where the authors performed a numerical study, relaxing the assumption of a frozen Higgs. They found the region of the parameter space that allows for the existence of the EWS, deriving an upper bound of approximately $8~\TeV$ for its mass, and they noted how the production and destruction of these states are $(B+L)$-violating processes. Other papers on the subject appeared
\cite{Ellis:2012cs,Kitano:2016ooc,Kitano:2017zqw,Hamada:2021oqm,Criado:2021tec}, discussing in particular the possibility of the EWS as a dark matter constituent.

The present work has the aim to progress along these lines by studying
some classical and quantum properties of the EWS.
In particular, we find that the mass of the EWS previously found in
the literature with a dynamical Higgs field was overestimated by an
order of magnitude.
We determine the rotational and vibrational modes of the EWS and use
this to argue, together with arguments on its topology and parity,
that the EWS is a spin-zero boson in its ground state.
We furthermore find results on its asymptotic interactions as well as
its interactions with fermions.
We finally discuss some constraints from phenomenology, which can
basically rule out the EWS in the semiclassical regime of parameter
space.

It is organized with the following structure.
In Section \ref{sec:modelews}, we review the EWS with dynamical Higgs and present our new numerical results.  
In Section \ref{sec:ewcollcoor}, we construct a Hilbert space for the
collective degrees of freedom of the Skyrmion, and find its spin,
statistics and other quantum numbers.
In Section \ref{sec:eqleptons}, we compute perturbatively the lepton number of the soliton.
In Section \ref{sec:longdistance}, we find an expression for the
long-distance interaction between two EWSs in the semiclassical
approximation. 
In Section \ref{sec:phenomenology}, some phenomenological aspects and
the EWS as a dark matter candidate are discussed. 
We conclude in Section \ref{sec:conclusion} with a discussion.
We have delegated details about the breather (or oscillon) to
App.~\ref{app:breather}, the asymptotic interaction potential to
App.~\ref{app:intpot} and the asymptotics of the fields to
App.~\ref{app:asymptotic_behavior}.

\section{The electroweak Skyrme model}
\label{sec:modelews}

The electroweak Skyrmion (EWS) is a soliton solution of the
higher-derivative extension of the electroweak (EW) theory, described
by the Lagrangian
\beq
\label{eq:ewextended}
\mathcal{L}_{\rm EWS}=\mathcal{L}_{\mathrm{EW}}+\frac{1}{8\Lambda^4}\mathrm{tr}\left((D_{\mu}\Phi)^{\dagger}(D_{\nu}\Phi)-(D_{\nu}\Phi)^{\dagger}(D_{\mu}\Phi)\right)^2 \ ,
\eeq
where $\mathcal{L}_{\rm EW}$ is the electroweak theory of Weinberg and
Salam with gauge group $SU(2)_L\times U(1)_Y$, which we will briefly
review:
\begin{equation}
    \mathcal{L}_{\rm EW}=-\frac{1}{2g^2}\mathrm{tr}\ W_{\mu\nu}W^{\mu\nu}-\frac{1}{2g'^2}\mathrm{tr}\ B_{\mu\nu}B^{\mu\nu}+\frac{1}{2}\mathrm{tr}\,(D_{\mu}\Phi)^{\dagger}(D^{\mu}\Phi)-\frac{\lambda}{4}\left(\mathrm{tr}(\Phi^{\dagger}\Phi)-v^2\right)^2\ ,
\end{equation}
where the $SU(2)_L$ gauge fields are
$W_{\mu}=gW_{\mu}^a\frac{\tau^a}{2}$, $a=1,2,3$, 
with field strength tensor
$W_{\mu\nu}=\partial_{\mu}W_{\nu}-\partial_{\nu}W_{\mu}-\i\left[W_{\mu}, W_{\nu}\right]$
and the $U(1)_Y$ gauge field is
$B_{\mu}=g'b_{\mu}\frac{\tau^3}{2}$,
with field strength tensor
$B_{\mu\nu}=\partial_{\mu}B_{\nu}-\partial_{\nu}B_{\mu}$.
The Higgs doublet is parametrized as a $2\times 2$ complex matrix with
real determinant
\begin{equation}
  \label{eq:higgsdoublet}
  \Phi=
  \begin{pmatrix}
    \i\tau^2\phi^* & \phi
  \end{pmatrix} = 
  \begin{pmatrix}
    \phi_0^* & \phi_1 \\
    -\phi_1^* & \phi_0
  \end{pmatrix}\ , \qquad
  D_{\mu}\Phi=\partial_{\mu}\Phi-\i W_{\mu}\Phi+\i\Phi B_{\mu} \ .
\end{equation}
With this parametrization, the $SU(2)_L$ group acts on the Higgs
doublet as a left multiplication and the hypercharge group, is a
gauged subgroup of $SU(2)_R$, acts as a right multiplication
\begin{align}
  \label{eq:su2l}
  SU(2)_L&: \Phi\longrightarrow L\Phi\ , \qquad W_{\mu}\longrightarrow LW_{\mu}L^{\dagger}-\i\partial_{\mu}LL^{\dagger} \ ,\\
  U(1)_Y&: \Phi\longrightarrow \Phi e^{-\i\varepsilon\frac{\tau^3}{2}}\ , \qquad B_{\mu}\longrightarrow B_{\mu}+\frac{\tau^3}{2}\partial_{\mu}\varepsilon \ ,
\end{align}
the Higgs sector manifestly enjoys a global $SU(2)_L\times SU(2)_R$
symmetry, which is commonly referred to as \emph{custodial symmetry}.
The Higgs doublet can also be parametrized as 
\begin{equation}
\label{eq:parametrization}
\Phi=sH \ , \qquad
s=\frac{v+h}{{\sqrt{2}}}\in \mathbb{R} \ ,\qquad
H\in SU(2) \ ,
\end{equation}
where $h$ describes the Higgs scalar, while the matrix $H$ describes
the would-be Goldstone bosons that originate by spontaneous symmetry
breaking. Passing to unitary gauge, one can eliminate $H$, the
would-be Goldstone bosons, leaving no further gauge freedom. Note that
the parametrization \eqref{eq:parametrization} is singular whenever
$s(x)=0$ at any $x\in\mathbb{R}^{1,3}$, which means that the unitary
gauge is generally ill-defined and fails to capture some
non-perturbative phenomena, e.g.~chiral anomalies.
The scalar field expectation value causes a spontaneous symmetry
breaking of the total gauge group $SU(2)_L\times U(1)_Y $ to its
diagonal subgroup $U(1)_{\rm em}$, which acts as 
\begin{equation}
	U(1)_{\mathrm{em}}: \Phi\longrightarrow e^{+\i\varepsilon\frac{\tau^3}{2}}\Phi e^{-\i\varepsilon\frac{\tau^3}{2}}\ , \qquad W_{\mu}\longrightarrow e^{+\i\varepsilon\frac{\tau^3}{2}}W_{\mu} e^{-\i\varepsilon\frac{\tau^3}{2}}+\frac{\tau^3}{2}\partial_{\mu}\varepsilon \ .
\end{equation}
The perturbative spectrum consists of the fields\footnote{We use the
convention in which an electrically charged elementary field with
charge $Q$
transforms as $\phi\mapsto e^{+\i\alpha Q}\phi$ under $U(1)_{\rm em}$.}
\begin{equation}
\label{eq:fields}
  A_{\mu}= \sin{\theta_w} W_{\mu}^3+\cos{\theta_w} B_{\mu}\ , \quad
  Z_{\mu}= \cos{\theta_w} W_{\mu}^3-\sin{\theta_w} B_{\mu} \ , \quad
  W_{\mu}^{\pm}=\frac{1}{\sqrt{2}}\left(W_{\mu}^1\pm \i W_{\mu}^2\right)\ ,
\end{equation}
where $\tan{\theta_w}=\frac{g'}{g}$ is the Weinberg angle.
$W_{\mu}^1$, $W_{\mu}^2$ and $Z_{\mu}$ acquire a mass
$m_{W}=m_Z \cos{\theta_w} =\frac{gv}{2}$,
while the remaining massless gauge boson $A_{\mu}$ is the gauge field
of $U(1)_{\rm em}$. The Higgs mass is
$m_h = \sqrt{2\lambda } v$.
The phenomenological values of the dimensionful parameters are
\beq
m_W = 80.379 \, \GeV\,, \qquad  m_h = 125.1 \, \GeV\,, \qquad v = 246 \, \GeV \ . 
\eeq
The numerical results in the rest of the paper are all done with these
values fixed.

The parametrization $\Phi$ of the Higgs doublet has the
advantage of providing an extremely simple form of the coupling to a
fermion doublet, whose mass matrix is proportional to the identity
\begin{equation}
  \label{eq:leptons}
  \mathcal{L}_{\mathrm{leptons}}=\bar{\psi}\i\slashed{D}\psi-y\bar{\psi}(\Phi P_R+\Phi^{\dagger}P_L)\psi\ ,
\end{equation}
where $y\langle\Phi\rangle=m_{\ell}\mathds{1}_2$ and the transformation laws of the fermions are
\begin{equation}
	\begin{split}
		L\in SU(2)_L: & \qquad  \psi_L\longrightarrow L\psi_L\ , \qquad \quad \psi_R\longrightarrow \psi_R \ ,\\
		e^{\i \alpha} \in U(1)_Y: & \qquad \psi_L\longrightarrow e^{\i Y \alpha}\psi_L\ , \qquad \psi_R\longrightarrow e^{\i Y \alpha+\i \alpha\frac{\tau^3}{2}}\psi_R\ ,
	\end{split}
\end{equation}
where $Y$ is the  hypercharge.

Now we turn to the analysis of the topological structure of the
classical field configuration space of the theory. We restrict to the
subspace of static field configurations with finite energy. The energy
functional of the EW part of the theory is
\begin{align}
  \label{eq:energy}
  E_{\rm EW}=\int\d^3x\ & \left[\frac{1}{g^2} \mathrm{tr}\ W_{0i}^2+\frac{1}{g'^2}\mathrm{tr}\ B_{0i}^2+\frac{1}{2} \mathrm{tr}\ (D_0\Phi)^{\dagger}(D_0\Phi) \right. \\
    &\left. \mathop+\frac{1}{2g^2} \mathrm{tr}\ W_{ij}^2+\frac{1}{2g'^2}\mathrm{tr}\ B_{ij}^2+\frac{1}{2} \mathrm{tr}\ (D_j\Phi)^{\dagger}(D_j\Phi)+\frac{\lambda}{4}\left(\mathrm{tr}(\Phi^{\dagger}\Phi)-v^2\right)^2 \right] . \nonumber
\end{align}
From this expression, it is evident that any static field
configuration with finite energy must approach to pure gauge at
spatial infinity
\begin{equation}
  W_{\mu}\underset{r\to\infty}\longrightarrow W_\mu^{\rm vac}=-\i H^{\dagger}\partial_{\mu}H, \qquad
  B_{\mu}\underset{r\to\infty}\longrightarrow B_\mu^{\rm vac}=\partial_{\mu}\varepsilon\frac{\tau^3}{2}, \qquad
  \Phi\underset{r\to\infty}\longrightarrow\Phi^{\rm vac}=\frac{v}{\sqrt{2}}H e^{-\i\varepsilon\frac{\tau^3}{2}} \ .\label{eq:clasvac}
\end{equation}
The topology of the static field configurations that satisfy these
boundary conditions can be partially classified through the winding
number of the $H$ field and with the Chern-Simons number of the weak
gauge field
\begin{align}
  \label{eq:winding}
  n_H&=-\frac{1}{24\pi^2}\varepsilon_{ijk}\int\d^3x\ \mathrm{tr}\left(H^{\dagger}\partial_i H\ H^{\dagger}\partial_j H\ H^{\dagger}\partial_k H \right) \ ,\\
  n_{\rm CS}&=-\frac{1}{16\pi^2}\varepsilon_{ijk}\int\d^3x\ \mathrm{tr}\left(W_iW_{jk}+\frac{2\i}{3}W_iW_jW_k\right) \ .
\end{align} 
The winding number $n_H$ is generally ill-defined. Due to the
singularity of the parametrization \eqref{eq:parametrization}, which
allows $H$ to be discontinuous whenever $s(x)=0$.
If, however $s(x)\neq 0$ everywhere, $n_H$ will be an integer thanks
to the boundary conditions \eqref{eq:boundary}
(i.e.~$\lim_{r\to\infty}H=\mathds{1}_2$).
Instead, $n_{\rm CS}$ is an integer only when $W_{\mu}=\i U^{\dagger}\partial_{\mu}U$,
($U\in SU(2)$) and in 
this case corresponds to the winding number of $U$.
$n_H$ and $n_{\rm   CS}$ are not gauge invariant. Under gauge
transformations with winding number $N$, they transform as
$n_H\rightarrow n_H+N$, $n_{\rm CS}\rightarrow n_{\rm CS}+N$.
Nevertheless, they can be used to construct the gauge invariant
quantity $n=n_H-n_{\rm CS}$, which elegantly can be written as
\begin{equation}
  \label{eq:topinv}
  n=-\frac{1}{24\pi^2}\varepsilon_{ijk}\int\d^3x\ \mathrm{tr}\left(H^{\dagger}D_iH\ H^{\dagger}D_jH\ H^{\dagger}D_kH+\frac{3\i}{2}H^{\dagger} W_{ij}D_kH\right) \ .
\end{equation}
For the classical vacuum field configurations given in
Eq.~\eqref{eq:clasvac}, $n_{\rm CS}=n_H=w(H)$.
Topologically distinct vacua are separated by energy barriers, whose
points of minimal height are unstable field configurations called
\emph{sphalerons} \cite{Klinkhamer:1984di}.
In this paper, we will use a ``modified Chern-Simons'' number
\cite{Ambjorn:1984bb,Criado:2020zwu}:
\beq
n_W = -\frac{1}{24\pi^2}\int d^3x\;\varepsilon_{ijk}\mathrm{tr}(\i W_iW_jW_k)\ ,
\eeq
which converges to the Chern-Simons number $n_{\rm CS}$ for pure gauge
configurations.

\subsection{The Skyrme term}

The addition of the Skyrme term to the electroweak Lagrangian has the aim to stabilize the Skyrmion in the limit in which the Higgs field and the weak gauge fields are frozen.
In this subsection, we show that the Skyrme term contains (almost) the only effective operators among those of dimension six and eight, that, at least in principle, meet this requirement. The operator we are looking for:
\begin{itemize}
\item[{\it i)}] must not disappear in the limit $g\rightarrow 0$, when the gauge fields are pure gauge,
\item[{\it ii)}]
  must not disappear in the limit $m_h\rightarrow \infty$, when the Higgs doublet is frozen,
\item[{\it iii)}] must scale at least as $\mu^{\alpha}$ with $\alpha<3$ under dilatations in order to circumvent Derrick's theorem.
\end{itemize}
At low energies, the physics beyond the Standard Model can be integrated out and its effects can be embedded in a momentum expansion in terms of non-renormalizable operators depending on the Standard Model fields
\begin{equation}
	\mathcal{L}=\mathcal{L}_{\mathrm{SM}}+\sum\frac{c_i^{d=6}}{\Lambda^2}\mathcal{O}_i^{d=6}+\sum\frac{c_i^{d=8}}{\Lambda^4}\mathcal{O}_i^{d=8}+\cdots
\end{equation}
There are two inequivalent ways of writing this series
\cite{Cohen:2020xca}. It can be expressed using the whole Higgs
doublet $\phi$ (see Eq.~\eqref{eq:higgsdoublet}) as a building block,
which leads to the Standard Model Effective Field Theory (SMEFT), or
it can be expressed treating $h$ and $H$ (see
Eq.~\eqref{eq:parametrization}) as unrelated fields, which leads to
the Higgs Effective Field Theory (HEFT).
In the rest of this work, we will adopt a SMEFT-based approach. 
Let us first look at the operators of dimension six, in the basis of Ref.~\cite{Murphy:2020rsh}:
\begin{center}
	\renewcommand{\arraystretch}{1.1}
	\begin{tabular}{|L|L||L|L|}
		\hline
		Q_{\phi W} & (\phi^{\dagger}\phi) W^{a}_{\mu\nu}W^{\mu\nu}_a & Q_W & \varepsilon_{abc}W^{a\nu}_{\mu}W^{b\rho}_{\nu}W^{c\mu}_{\rho} \\ 
		
		Q_{\phi \widetilde{W}} & (\phi^{\dagger}\phi) W^{a}_{\mu\nu}\widetilde{W}^{\mu\nu}_a & Q_{\widetilde{W}} & \varepsilon_{abc}\widetilde{W}^{a\nu}_{\mu}W^{b\rho}_{\nu}W^{c\mu}_{\rho} \\ 
		
		Q_{\phi B} & (\phi^{\dagger}\phi)B_{\mu\nu}B^{\mu\nu} & Q_{\phi} & (\phi^{\dagger}\phi)^3 \\ 
		
		Q_{\phi \widetilde{B}} & (\phi^{\dagger}\phi)B_{\mu\nu}\widetilde{B}^{\mu\nu} & Q_{\phi \square} & (\phi^{\dagger}\phi)\square (\phi^{\dagger}\phi) \\ 
		
		Q_{\phi WB} & (\phi^{\dagger}\tau^a\phi) \widetilde{W}^{a}_{\mu\nu}B^{\mu\nu} & Q_{\phi D} & (\phi^{\dagger}D^{\mu}\phi)^*(\phi^{\dagger}D_{\mu}\phi) \\ 
		
		Q_{\phi \widetilde{W}B} & (\phi^{\dagger}\tau^a\phi) \widetilde{W}^{a}_{\mu\nu}B^{\ \mu\nu} & & \\ 
		\hline 
	\end{tabular} 
	\captionof{table}{Dimension-6 operators in the SMEFT Lagrangian involving electroweak fields.}
\end{center}
These operators cannot stabilize the Skyrmion. All the operators that 
contain a field strength tensor vanish when the fields are pure
gauge. The remaining operators can be excluded due to Derrick's
theorem.

We are forced to search for a suitable extension among the dimension-8
operators; the following table presents the operators that do not
involve field strength tensors:
\begin{center}
	\renewcommand{\arraystretch}{1.1}
	\begin{tabular}{|L|L||L|L|}
		\hline
		Q_{\phi^8} & (\phi^{\dagger}\phi)^4 & Q_{\phi^4}^{(1)} & (D_{\mu}\phi^{\dagger}D_{\nu}\phi)(D^{\nu}\phi^{\dagger}D^{\mu}\phi) \\
		Q_{\phi^6}^{(1)} & (\phi^{\dagger}\phi)^2(D_{\mu}\phi^{\dagger}D^{\mu}\phi) & Q_{\phi^4}^{(2)} & (D_{\mu}\phi^{\dagger}D_{\nu}\phi)(D^{\mu}\phi^{\dagger}D^{\nu}\phi) \\
		Q_{\phi^6}^{(2)} & (\phi^{\dagger}\phi)(\phi^{\dagger}\tau^a\phi)(D_{\mu}\phi^{\dagger}\tau^aD^{\mu}\phi) & Q_{\phi^4}^{(3)} & (D^{\mu}\phi^{\dagger}D_{\mu}\phi)(D^{\nu}\phi^{\dagger}D_{\nu}\phi) \\
		\hline
	\end{tabular}
	\captionof{table}{Dimension-8 operators in the SMEFT Lagrangian involving electroweak fields.}
\end{center}
As a consequence of Derrick's theorem, the only operators that can
stabilize the Skyrmion are $Q_{\phi^4}^{(1)}$, $Q_{\phi^4}^{(2)}$ and
$Q_{\phi^4}^{(3)}$. The Skyrme term \eqref{eq:ewextended} corresponds
to the particular combination of operators:
\begin{equation}
	\mathcal{O}_{\rm Sk}=\frac{1}{2}\big(Q_{\phi^4}^{(1)}+Q_{\phi^4}^{(2)}-2Q_{\phi^4}^{(3)}\big) \ .
\label{skterm}
\end{equation}
Custodial symmetry is satisfied as long as the coefficients of
$Q_{\phi^4}^{(1)}$ and $Q_{\phi^4}^{(2)}$ are equal. In principle, by
choosing different coefficients of the three operators, or by adding
other operators of dimensions six and eight, it is still possible to
obtain a stable soliton, and it is expected that these operators give
non-negligible contributions to the EWS computation.
These possibilities have been studied in
Ref.~\cite{Criado:2021tec}, in the HEFT context, but will not be taken
into account in the present work. Our choice has the advantage of
yielding equations of motion that are linear in second-derivatives of
the fields, as pointed out in Ref.~\cite{Adkins:1983ya}. We will not
discuss the origin of the Skyrme term, which can be the low-energy
counterpart of a variety of renormalizable extensions of the Standard
Model \cite{Criado:2020zwu}, in particular of those including massive
fermions \cite{DHoker:1984mif, DHoker:1984izu, DasBakshi:2020ejz}.

\subsection{The electroweak Skyrmion solution}
\label{sec:asymp}

The Ansatz for the weak gauge field configuration is chosen to be spherically symmetric\footnote{Minimizing over the field configurations of this form yields a true minimum of the potential thanks to the principle of symmetric criticality \cite{Manton:2004tk}.}
\begin{equation}
	\label{eq:profile}
	W_i(\textbf{x})=g\frac{\tau^a}{2}\left[\varepsilon_{aij}\hat{x}_j\frac{f_1(r)}{r}+(\delta_{ia}-\hat{x}_i\hat{x}_a)\frac{f_2(r)}{r}+\hat{x}_i\hat{x}_a\frac{b(r)}{r}\right], \qquad W_0(\textbf{x})=0 \ .
\end{equation}
Field configurations of this kind have the nice property of being invariant under combined global $SU(2)_L$ and spatial $SO(3)$ transformations
\begin{equation}
	\label{eq:rotations}
	W_i^{(D)}(\textbf{x})=D_{ji}W_j(D\textbf{x})=AW_i(\textbf{x})A^{\dagger} \ , \qquad D\in SO(3)\ , \  A\in SU(2)
\end{equation}
The Higgs field configuration is chosen to be of the form
\begin{equation}
\label{eq:higgs}
	\Phi(\textbf{x})=\frac{v\sigma(r)}{\sqrt{2}}\mathds{1}_{2}\ ,
\end{equation}
which is invariant under spatial rotations and under
$SU(2)_{\mathrm{diag}}\subset SU(2)_L\times SU(2)_R$ transformations. 
Therefore, in this gauge the solution has zero Higgs winding number,
and is characterized by its Chern-Simons number only.

We briefly review some previous results on this topic.
The solutions have been studied in the literature in the limit where the Higgs field \cite{Eilam:1985tg,Ambjorn:1984bb,Farhi:1995aq} or the gauge fields \cite{Kitano:2016ooc} are frozen. When both limits are taken $(g\rightarrow 0, m_h\rightarrow\infty)$, the Lagrangian takes the form
\begin{equation}
	\mathcal{L}=\frac{v^2}{4}\mathrm{tr}\ (\p_{\mu}H)^{\dagger}(\p^{\mu}H)+\frac{v^4}{32\Lambda^4}\mathrm{tr}\left[H^{\dagger}\partial_{\mu}H, H^{\dagger}\partial_{\nu}H\right]^2,
\end{equation}
which is a scaled-up version of the Skyrme Lagrangian with massless mesons. In this reduced model, the Skyrmion mass would be
\beq
M_{\rm EWS} \simeq 23.2 \, \pi \frac{v^3}{\Lambda^2}  \approx  \frac{10^5 \ \GeV }{(\Lambda/100 \ \GeV)^2}\;,  \label{skestimate}
\eeq
and its lifetime infinite. 
In Ref.~\cite{Ambjorn:1984bb} the authors work under assumption of non-dynamical Higgs and dynamical gauge fields ($g\neq 0, m_h\to\infty$). The results are expressed as a function of the dimensionless parameter $\xi=\frac{4\Lambda^4}{g^2v^4}$. It was found that the Skyrmion has a
mass \cite{Ambjorn:1984bb}
\begin{equation}
    \xi_{\rm crit} \simeq 10\ , \qquad \Lambda_{\rm crit} \simeq 250 \ \GeV\ , \qquad M_{\rm EWS} \simeq 15 \ \TeV \ .
\end{equation}
where $\xi_{\rm crit}$ and $\Lambda_{\rm crit}$ are the lowest values
of the parameter $\xi$ and $\Lambda$ for which the EWS is classically
stable (see Sec.~\ref{sec:stability} for a more detailed
discussion). In Ref.~\cite{Criado:2020zwu} the EWS is studied using
the most general assumptions of dynamical Higgs and gauge fields
($g\neq0, m_h\neq\infty$).
The shape of the Skyrmion was obtained as a function of the
modified Chern-Simons number $n_W$ of the field configuration at fixed
$\Lambda$, by introducing the Lagrange multiplier $\lambda$ and
minimizing the functional:
\begin{equation}
  E_{\lambda}=E-\lambda\big(n_W-n_W^{(\rm Sk)}\big)\ .
  \label{eq:Elambda}
\end{equation}
In this way, one can obtain the energy of the EWS (if
the solution exists) as a function of $\Lambda$ in parameter space,
and of $n_W$ in field configuration space.  In
Ref.~\cite{Criado:2020zwu} the Skyrmion mass and radius relations were
obtained:
\begin{equation}
  \label{eq:massradius}
  M\simeq 0.35\frac{4\pi v^3}{\Lambda^2}\ , \qquad R\simeq 0.6\frac{v}{\Lambda^2}\ ,
\end{equation}
and the mass is minimized quite close to, but not exactly at $n_W=1$.
All current literature  considers the $U(1)_Y$
fields decoupled ($g'=0, B_{\mu}=0$) at the zeroth order of the
semiclassical approximation. This assumption is reasonable as long as
$g'\ll g$, which is true in nature as
$\sin^2{\theta_{\mathrm{W}}}\simeq 0.223$.

\subsection{Numerical results}\label{sec:numresults}

The shape of the EWS can be determined by minimizing the integral
obtained by substituting the spherically symmetric Ansatz \eqref{eq:profile} into the
static energy functional of the theory \eqref{eq:ewextended}. Thanks
to spherical symmetry, the resulting integral is one dimensional, and
the integrand is a polynomial in  $f_1(r), f_2(r), b(r)$, $\sigma(r)$
and their first derivatives:
\beq
\label{eq:eenat}
E = \frac{4\pi v^3}{\Lambda^2}E_{\rm nat}\ ,
\eeq
with the dimensionless energy functional\footnote{We note that there were two typos in the reduced energy in Ref.~\cite{Criado:2020zwu}, i.e.~a factor of $\sigma^2$ was missing in the last term and the sign in front of $(2f_1-1)b/r$ was opposite in the second term.}
\begin{align}
\label{eq:enat}
  E_{\rm nat}=\int\d r&\Bigg\{\alpha\left[\left(f_1'-2f_2\frac{b}{r}\right)^2+\left(f_2'+(2f_1-1)\frac{b}{r}\right)^2+\frac{2}{r^2}\left(f_1^2+f_2^2-f_1\right)^2\right] \non
  &+ \frac{r^2}{2}(\sigma')^2 + \sigma^2\left(f_1^2+f_2^2+\frac{b^2}{2}\right)+\beta r^2(\sigma^2-1)^2 \non
  &+ (f_1^2+f_2^2)\sigma^2\left[(\sigma')^2+\frac{\sigma^2}{r^2}\left(b^2+\frac{f_1^2+f_2^2}{2}\right)\right] \Bigg\},
\end{align}
where we have performed the rescaling \cite{Criado:2020zwu} of the
radial coordinate by $r\to\frac{v}{\Lambda^2}r$ and the fields by
$(f_1,f_2,b)\to\frac2g(f_1,f_2,b)$ and we have defined
\begin{equation}
\label{eq:alphabeta}
  \alpha = \frac{\Lambda^4}{v^2m_W^2}\ , \qquad
  \beta = \frac{m_h^2 v^2}{8\Lambda^4}\ .
\end{equation}
Note that in this functional there are no derivatives of $b$, which
therefore is not a proper dynamical variable but a Lagrange multiplier
and can be solved algebraically
\beq
b = \frac{2r\alpha(2f_2f_1' + (1 - 2f_1)f_2')}{2\alpha((1-2f_1)^2 + 4f_2^2) + r^2\sigma^2 + 
  2(f_1^2+f_2^2)\sigma^4}\ .
\label{eq:bsol}
\eeq
The corresponding Euler-Lagrange equations
\begin{align}
  f_i'' - \frac{1}{8\pi\alpha}\frac{\p E_{\rm nat}}{\p f_i}
  &= \ddot{f}_i\ , \qquad i=1,2,\ \label{eq:eom_f}\\
  \sigma'' - \frac{1}{4\pi(r^2 + 2(f_1^2+f_2^2)\sigma^2)}\frac{\p E_{\rm nat}}{\p \sigma}
  &= \ddot{\sigma}\ ,\label{eq:eom_sigma}
\end{align}
must be solved numerically with the boundary conditions
\begin{equation}
  \label{eq:boundary}
  \begin{split}
    f_1(0)=&f_1'(0)=f_2(0)=b(0)=f_2'(0)-b'(0)=\sigma'(0)=0 \ , \\
    &f_1(\infty)=f_2(\infty)=b(\infty)=\sigma(\infty)-1=0 \ ,
  \end{split}
\end{equation}
which ensure the finiteness of the energy and the regularity at the
origin.

The modified Chern-Simons number for this configuration in dimensionless
coordinates is  
\begin{equation}
n_W=\frac{2}{\pi}\int\frac{1}{r}\d r\ (f_1^2+f_2^2)b\ .
\label{eq:csnumber_dimless}
\end{equation} 

In order to solve the field equations, mathematically we could
eliminate $b$ algebraically, but it turns out not to be the best way
for the numerical methods we have utilized. Our numerical method is
the following. 
We solve the second-order differential equations
\eqref{eq:eom_f}-\eqref{eq:eom_sigma} using the arrested Newton-flow
and updating $b$ by its exact solution \eqref{eq:bsol} at every
step. The arrested Newton-flow algorithm works by evolving the flow
``time'' and evaluating the potential energy
$E_{\rm nat}[f_1,f_2,\sigma]$ at every step, setting the first flow-time
derivative to zero ($\dot{f}_1=\dot{f}_2=\dot\sigma=0$) if $E_{\rm nat}$
increases.
At every step, $b$ is updated using its exact solution \eqref{eq:bsol}.
We measure the right-hand side of the equations and
after they integrate to a number smaller than $10^{-6}$, we stop the
flow and the numerical solution has been obtained. Contrary to claims
in the literature, we do not need to restrict the fields to a specific
Chern-Simons number if our initial guess is close enough to the
correct solution.

In order to define the size or radius of the EWS, we define the
following mean-squared functional 
\beq
\label{eq:rx}
r_X^2 := \frac{\int\d r\ r^4 X}{\int\d r\ r^2 X}\ ,
\label{eq:radius_def}
\eeq
where we can take $X$ to be the energy density $X=\mathcal{E}$, the
Chern-Simons density $X=\mathcal{Q}_W$ or simply $X=(1-\sigma)^2$.

\begin{figure}[!ht]
  \centering
  \includegraphics[width=0.49\textwidth]{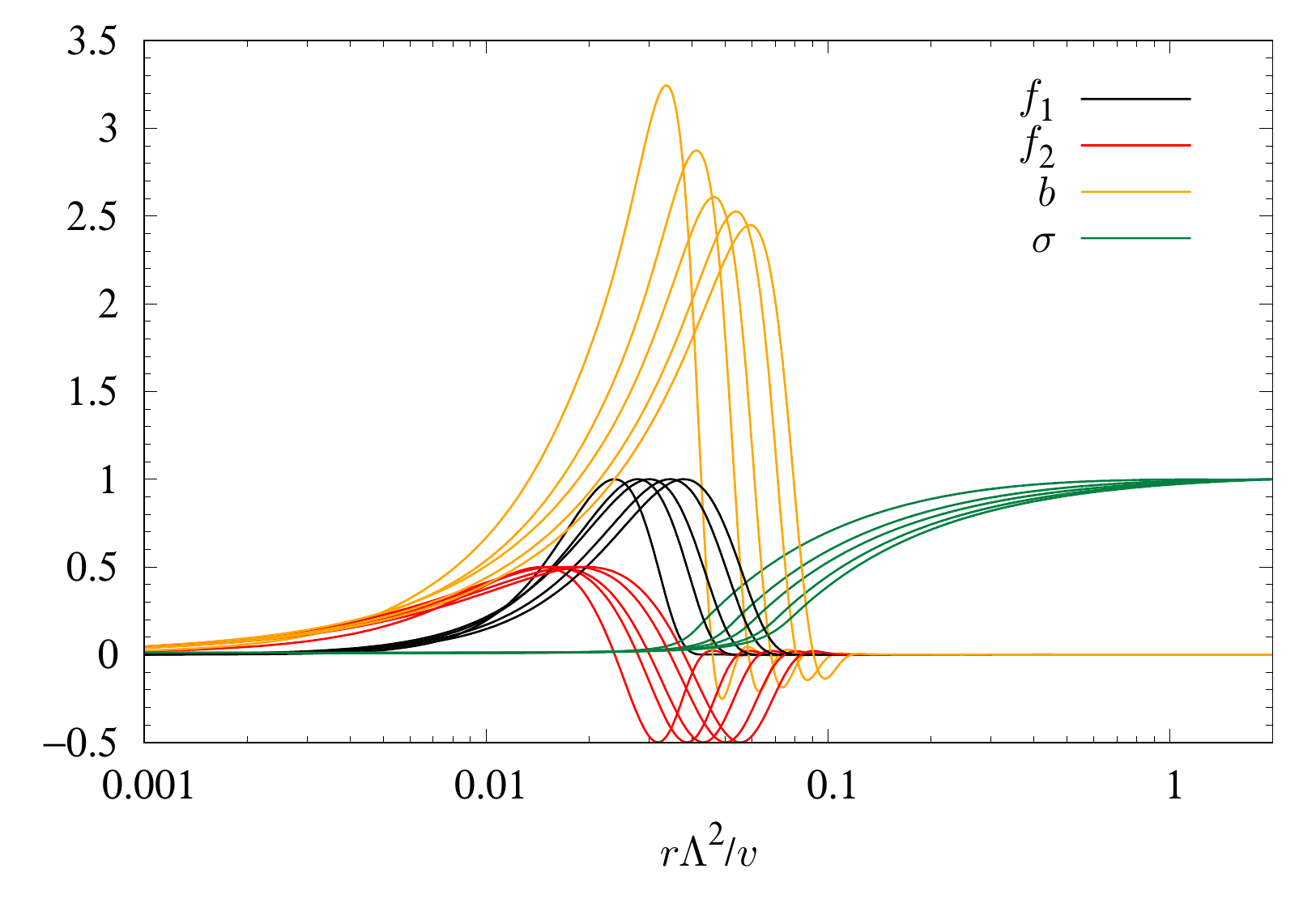}
  \includegraphics[width=0.49\textwidth]{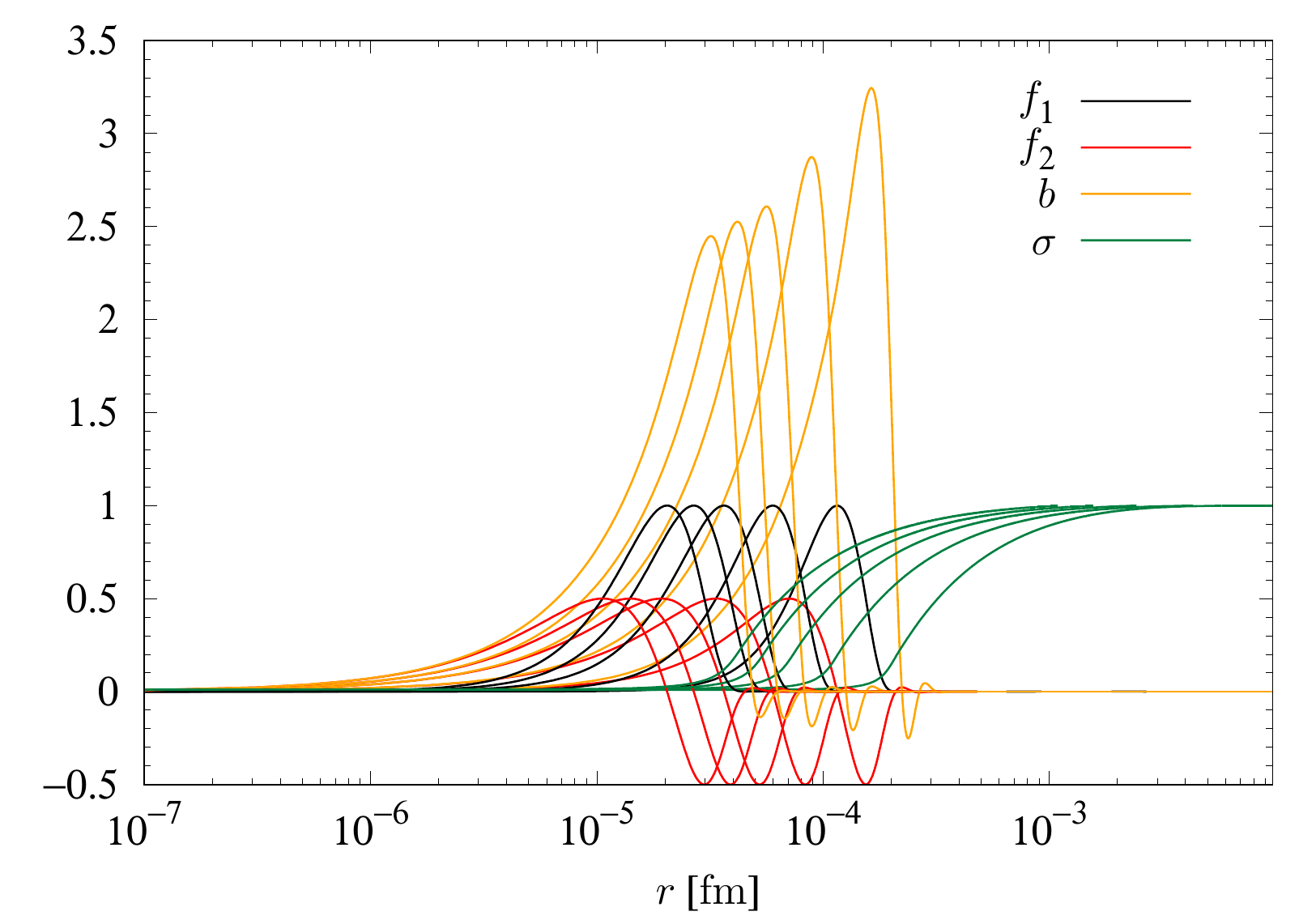}
  \caption{\footnotesize
    Electroweak Skyrmion profile functions for different values of
    $\Lambda=100,150,200,250,300$ GeV, corresponding to the curves
    from left to right (right to left) in the left (right) figure,
    plotted as functions of the log radius $r\Lambda^2/v$ ($r$).
    The modified Chern-Simons charge is localized where the Higgs
    field vanishes. }
    \label{fig:sols}
\end{figure}
\begin{figure}[!ht]
    \centering
   \includegraphics[width=0.49\textwidth]{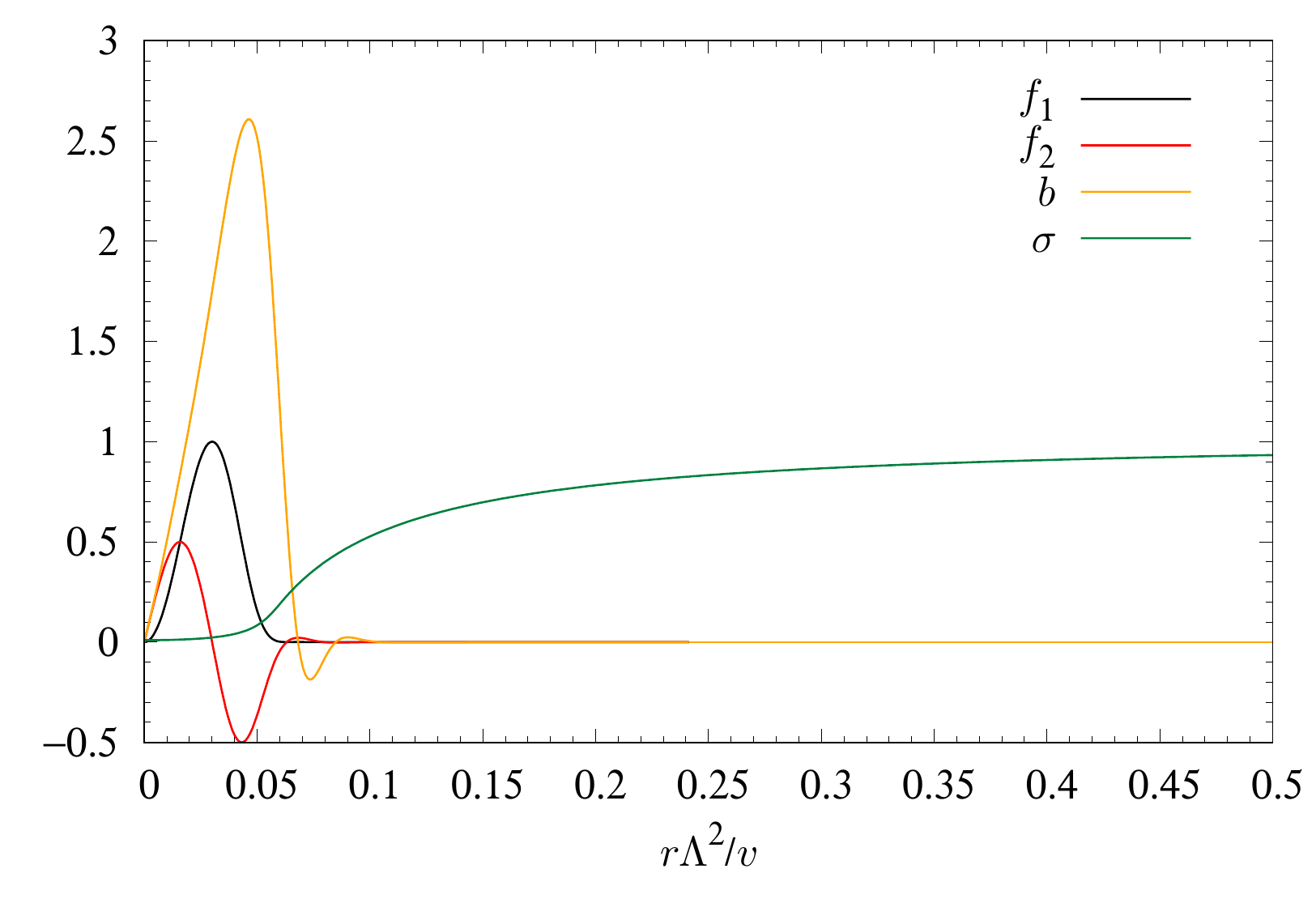}
   \includegraphics[width=0.49\textwidth]{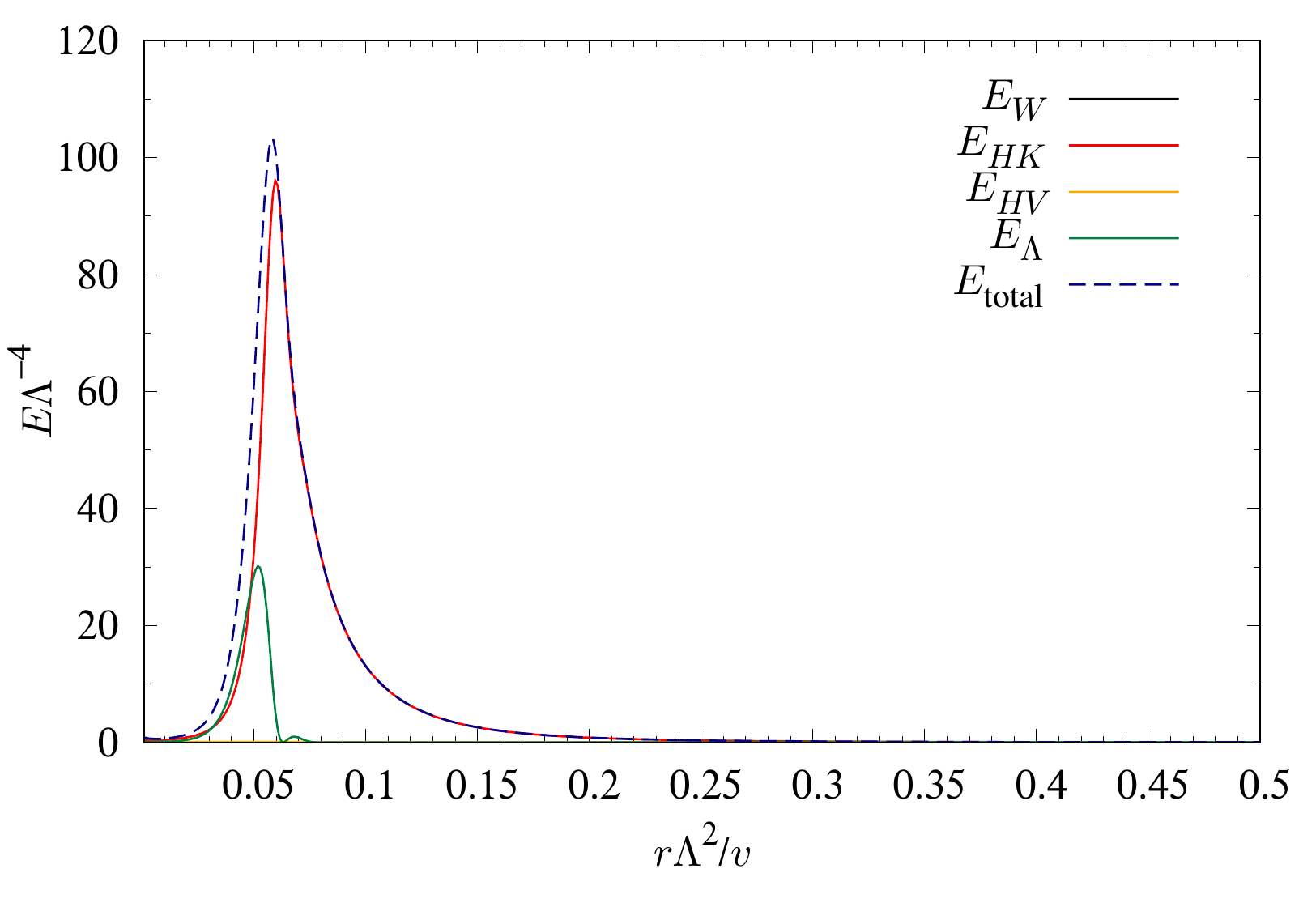}
   \caption{\footnotesize
     Electroweak Skyrmion profile functions (left) and energy
     densities (right) for $\Lambda=200$ GeV. }
    \label{fig:sol200}
\end{figure}
\begin{figure}[!ht]
  \centering
  \mbox{\subfloat[]{\includegraphics[width=0.49\textwidth]{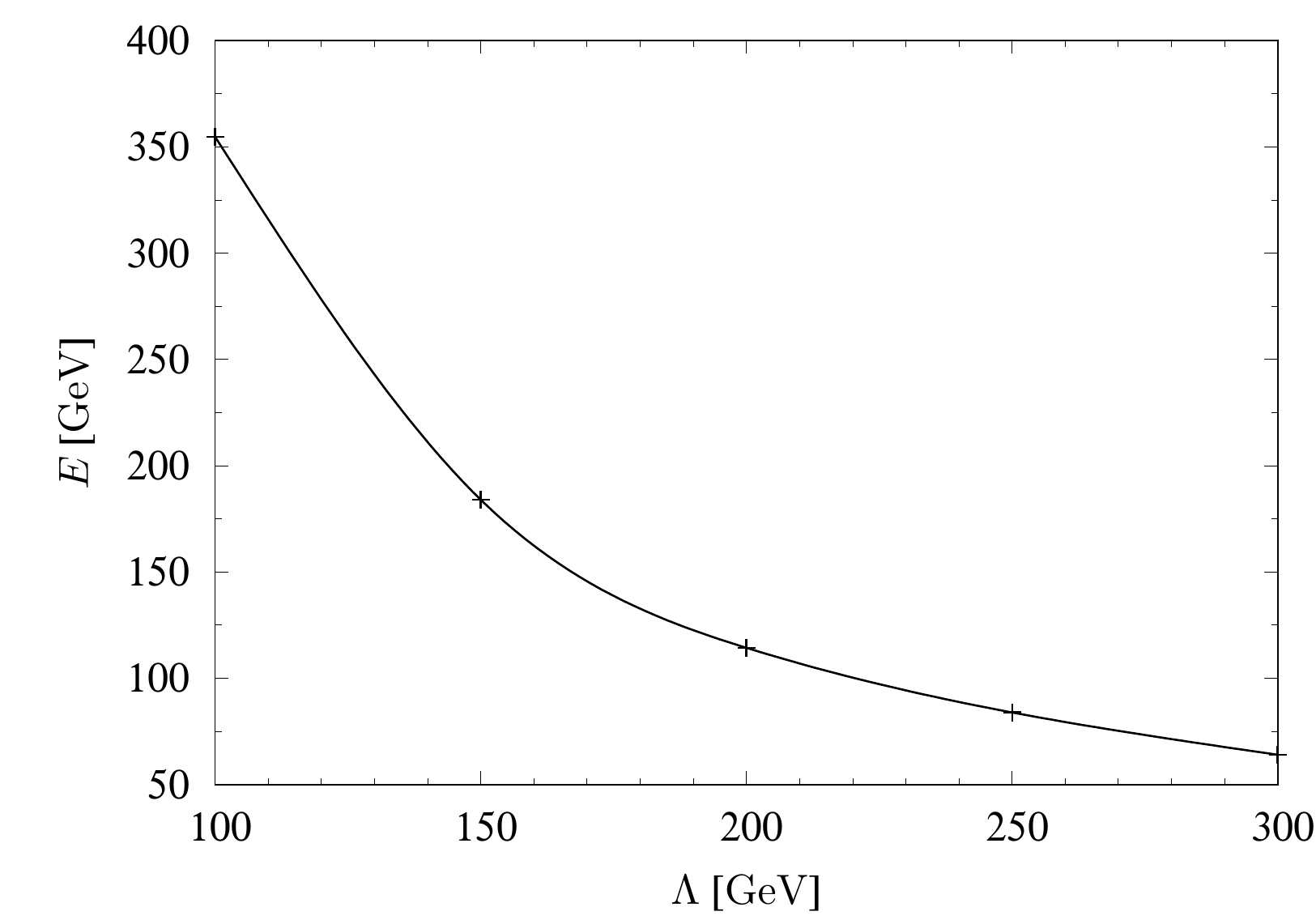}}
    \subfloat[]{\includegraphics[width=0.49\textwidth]{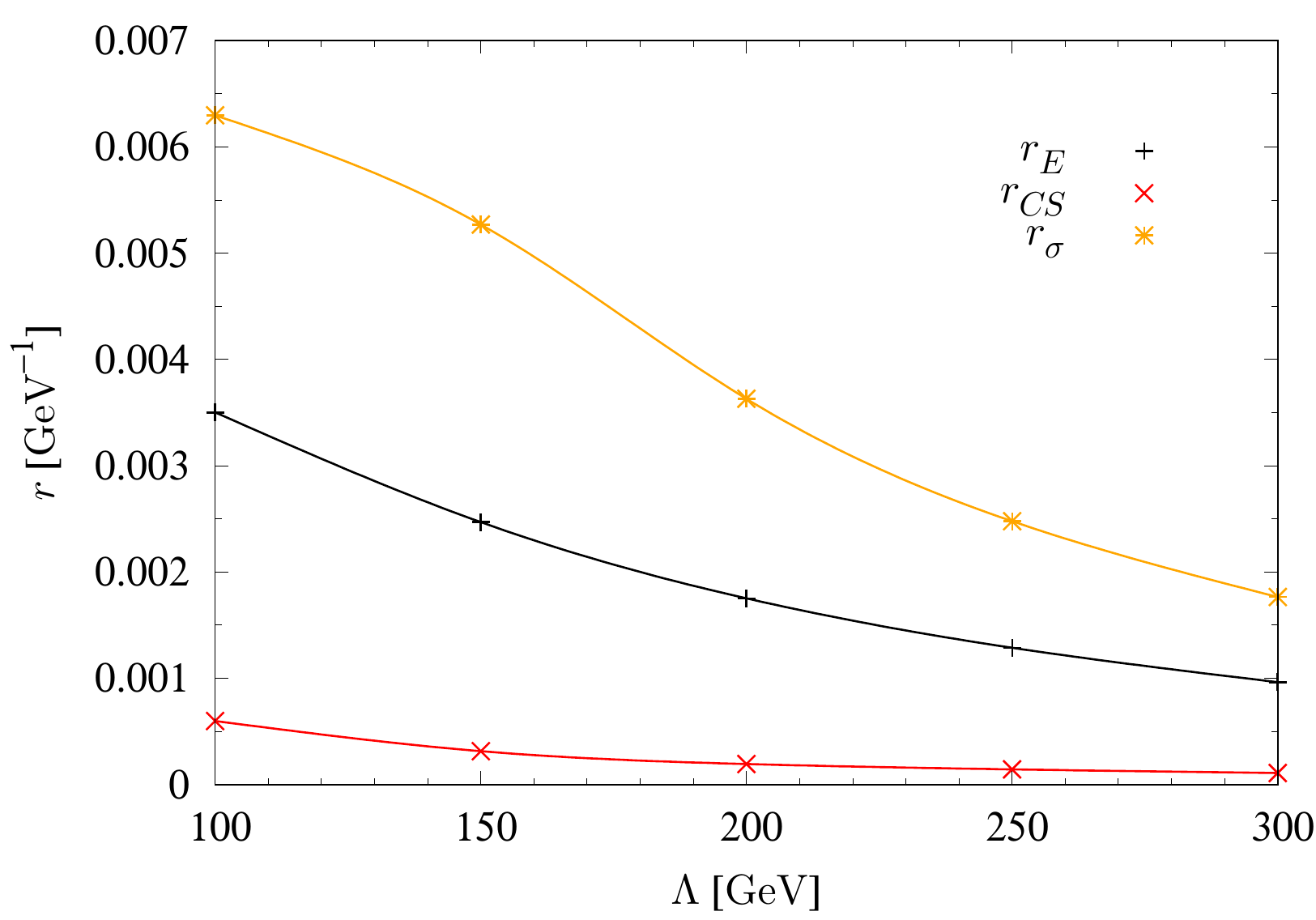}}}
  \caption{\footnotesize
    Mass (left) and radius (right) in GeV (GeV$^{-1}$) for the
    different EWSs of Fig.~\ref{fig:sols}.
    Three different definitions of the radius are shown, where
    Eq.~\eqref{eq:radius_def} is used with the energy density, the
    modified Chern-Simons density and $(1-\sigma)^2$, respectively. }
  \label{fig:energy_radii}
\end{figure}
\begin{figure}[!ht]
  \centering
  \mbox{\subfloat[]{\includegraphics[width=0.49\textwidth]{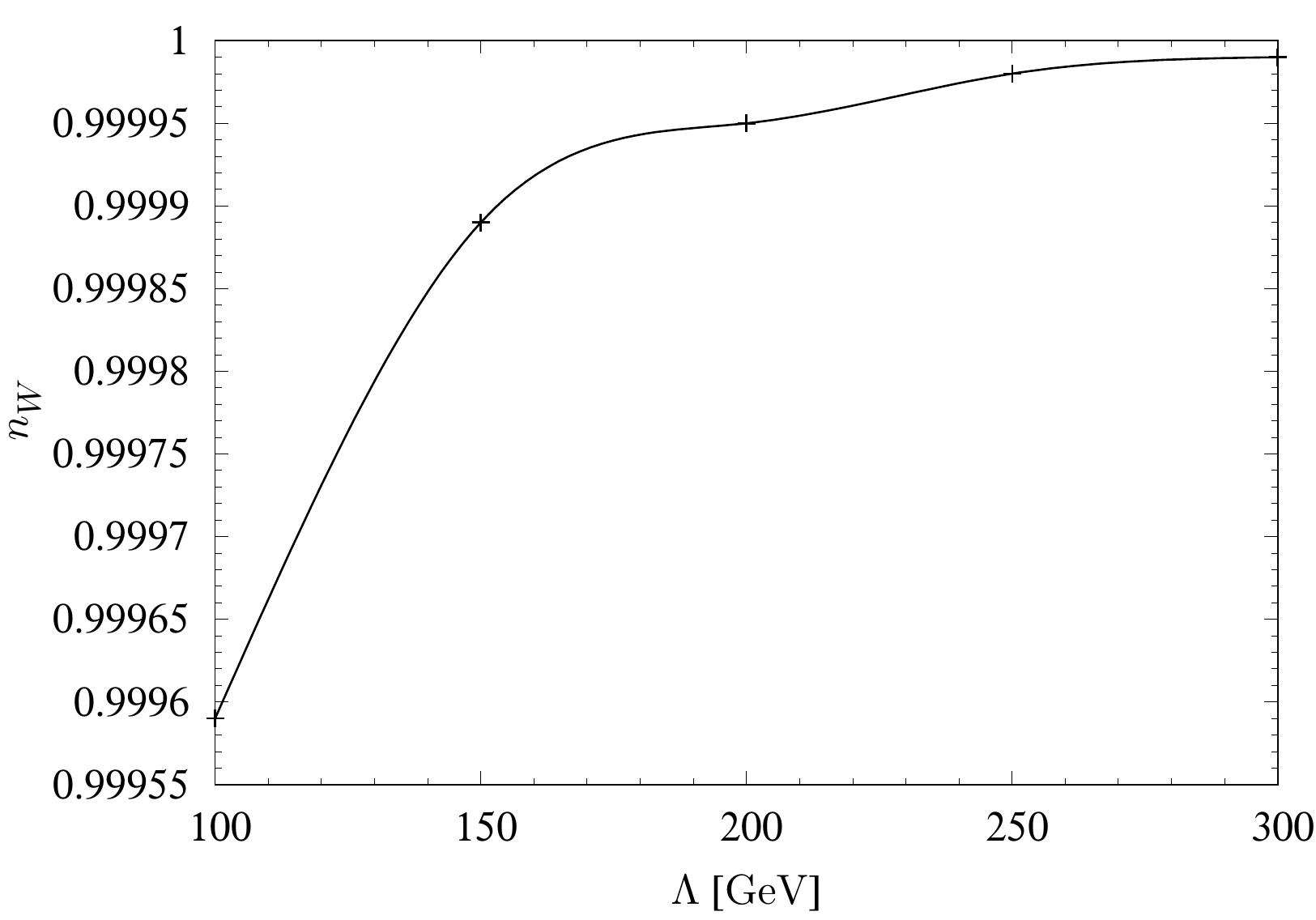}}
    \subfloat[]{\includegraphics[width=0.49\textwidth]{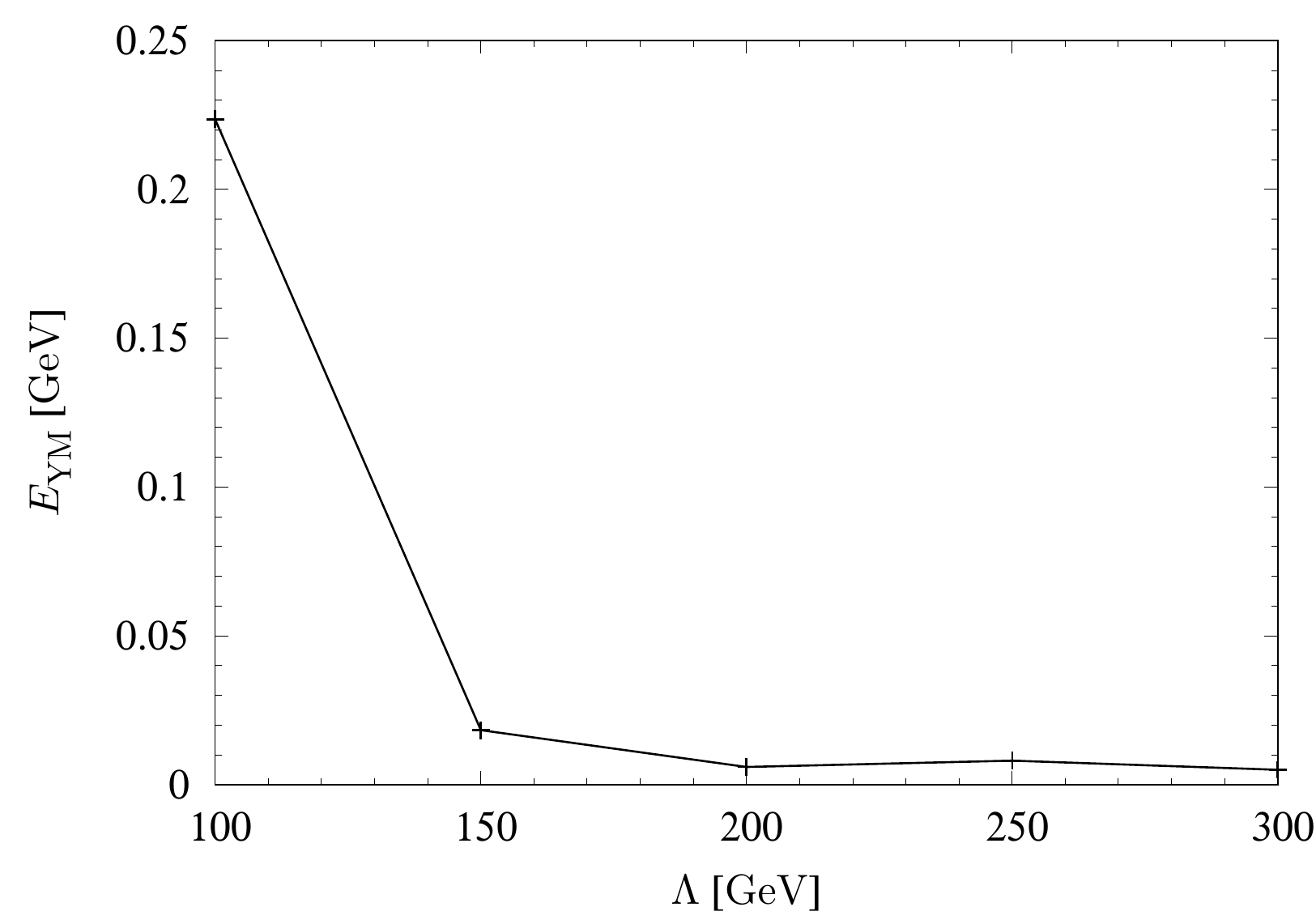}}}
  \caption{\footnotesize
    (a) Modified Chern-Simons number \eqref{eq:csnumber_dimless} for
    the solutions of Fig.~\ref{fig:sols}.
    They are so close to one for the entire range of $\Lambda$, that
    we cannot determine whether they are not exactly equal to one from
    their numerical error.
    (b) Energy of the YM tensor in GeV for the same solutions.
  } 
  \label{fig:cs}
\end{figure}
We now present our numerical results.
In Fig.~\ref{fig:sols} we show the profile functions $f_1$, $f_2$,
$b$, and $\sigma$ for various values of $\Lambda$ in the range
$\Lambda\in[100,300]~\GeV$.
The smaller $\Lambda$ is, the smaller is the size of the solution in
dimensionless coordinates.
This is an artifact of the dimensionless units, because in physical
units when restoring $v/\Lambda^2$, the solutions do shrink as
$\Lambda$ is increased. 
In Fig.~\ref{fig:sol200}(a) the profile functions are shown in a
non-log scale plot for $\Lambda=200~\GeV$, whereas in
Fig.~\ref{fig:sol200}(b) the corresponding energy density is shown.
In Fig.~\ref{fig:energy_radii}(a), the energy of the EWS is shown in
GeV as a function of $\Lambda$.
The energy or mass of the EWS clearly decreases as $\Lambda$
increases, as one would expect.

In Fig.~\ref{fig:energy_radii}(b), the radius of the EWS is shown in
fm according to various definitions of the density used as a measure
and for the range of $\Lambda$ used for the solutions in
Fig.~\ref{fig:sols}.
The radius of the EWS lies in the $10^{-3}$-$10^{-4}$ fm range and
shrinks as the scale $\Lambda$ is increased.
We can also see that the energy density is about an order of magnitude
more extended than the modified Chern-Simons density; this is largely due to
the tail of the Higgs field.
Approximate relations of the energy and radius obtained with the
energy density as functions of $\Lambda$ and $v$ are given by
\begin{align}
M&\simeq\left(0.0130 + \frac{\Lambda}{16.9\ \TeV}\right)
\frac{4\pi v^3}{\Lambda^2}\ ,\non
r_E&\simeq\left(-0.0749 + \frac{\Lambda}{390\ \GeV} - \frac{\Lambda^2}{(513\ \GeV)^2}\right)
\frac{v}{\Lambda^2}\ .
\label{fitsmrl}
\end{align}
Finally, in Fig.~\ref{fig:cs}(a) we show the modified Chern-Simons number
\eqref{eq:csnumber_dimless} for the solutions of
Fig.~\ref{fig:sols}. This number is always very close to one.
Since the number is so close to unity, we cannot determine whether it
is smaller than one or that is simply the numerical error of the
solution.
In Fig.~\ref{fig:cs}(b), we display the energy of the YM field
strength tensor, which can be seen to be four orders of magnitude
smaller than the mass of the EWS for $\Lambda\geq150\,\GeV$.
This means that the gauge field $W_i$ is very close to being pure
gauge, which explains why $n_W$ is so close to unity.

We find here that in Ref.~\cite{Criado:2020zwu} both the mass and the
radius are overestimated by more than an order of magnitude, probably
due to the numerical method utilized in that work.

\subsection{Quantum and higher-derivative corrections}
\label{sec:qcorr}

An important comment in store is about quantum corrections.  
In Fig.~\ref{fig:energy_radii}(a), we notice that around
$\Lambda\approx200~\GeV$, the mass of the EWS becomes of the order of
the Higgs mass, and thus of the scale of the most massive perturbative
particle. We can thus say that at this scale there is a threshold for
the soliton transitioning from a (semi-)classical object to a quantum
soliton. Another way to infer this is to estimate when the Compton
wave length of the soliton $\lambda_{\mathrm{Compton}}\sim\frac{1}{M}$
becomes of the same order of the soliton size.
Using the relations \eqref{fitsmrl} we have $\frac{1}{M}\sim  R$ for
$\Lambda\simeq 130\ \GeV$. 

The Lagrangian \eqref{eq:ewextended} defines an effective field theory
with a physical cutoff $\Lambda$. In order for our treatment to be
consistent, we need to have $R\gg \frac{1}{\Lambda}$. The $r_{\sigma}$
notion in Fig.~\ref{fig:energy_radii}(b), which is the most adequate
to compare higher-derivative corrections, is consistently of the same
order of $\frac{1}{\Lambda}$. 
Therefore, we must conclude that for these values of $\Lambda$, a fully
consistent treatment of the EWS requires a more
detailed knowledge of the UV completion of the SMEFT Lagrangian.
This is very similar to what happens for the Skyrmion in QCD with
higher-derivative corrections. Corrections are expected to be
important and of order one, but still under control.

\subsection{Stability of the EWS}
\label{sec:stability}
A soliton is \emph{classically} stable if it is a local
minimum of the energy in field configuration space. The classical
stability of the EWS depends on the value of the parameter
$\Lambda$. We briefly review the previous results on this topic. As we
already observed, in the limit $g\to0, m_h\to\infty$, the EWS
corresponds to a rescaled version of the QCD Skyrmion, and hence is
completely stable.
In Ref.~\cite{Ambjorn:1984bb}, under the assumption of non-dynamical
Higgs ($g\neq 0, m_h\to\infty $) the classical stability criterion of
the Skyrmion is found to be
\begin{equation}
  \xi\geq \xi_{\rm crit}\simeq 10.35\ , \qquad
  M\lesssim 15\ \TeV\ .
\end{equation}
where $\xi=\frac{4\Lambda^4}{g^2v^4}$.
On the other hand, using a dynamical Higgs field,
Ref.~\cite{Criado:2020zwu} found that classically stable solutions
exist only for 
$\Lambda\geq \Lambda_{\rm crit}\simeq 90\ \mathrm{GeV}$, which leads
to the following upper bound on the mass 
\begin{equation}
  M\lesssim 8\ \TeV\ .
  \label{eq:Criado_bound}
\end{equation}

The decay of the EWS can also happen through quantum effects: if the
EWS has no topological quantum number (which is the case) it can decay
by tunneling through the barrier that separates it from the
perturbative vacuum. There are at least two decay channels. Let us
consider for simplicity a solitonic field configuration with
$(n_W,n_H)\simeq(1,0)$. We have the following possibilities 
\begin{itemize} 
\item The EWS can tunnel to the vacuum with
  $(n_W,n_H)_{\mathrm{vac}}=(0,0)$ with a change of the (modified) Chern-Simons 
  number only. 
\item The EWS can tunnel to the vacuum with
  $(n_W,n_H)_{\mathrm{vac}}=(1,1)$ with a change of the Higgs
  winding number only. 
\end{itemize}
These two processes are not equivalent, and, in general, happen at
different rates. This can be seen by taking the limit,
$m_h\rightarrow\infty$, in which the Higgs field is frozen at its
vacuum expectation value: if we increase $m_h$, the fluctuations of
the Higgs field are more and more suppressed, and the energy barrier
between two field configurations that differ by one unit of $n_H$ only
becomes higher and higher. When $m_h$ is infinite, $n_H$ becomes a
topological invariant and the above-mentioned energy barrier becomes
infinite, making tunneling in that direction impossible. An estimate
of the decay rate of the EWS has been made in
Ref.~\cite{Rubakov:1985it} under the assumption of a frozen Higgs
field. As the discussion above implies, this estimate completely
misses decay processes of the second kind. The dynamical Higgs has two
competing effects on the stability of the EWS. On one hand it lowers
its mass rendering it more stable. On the other hand it opens new decay
possibilities, in other words it lowers also the potential barrier. To
compute this properly, at least in the semiclassical limit, it is
necessary to find the bounce solution.

\begin{figure}[!htp]
  \begin{center}
    \includegraphics[width=0.5\linewidth]{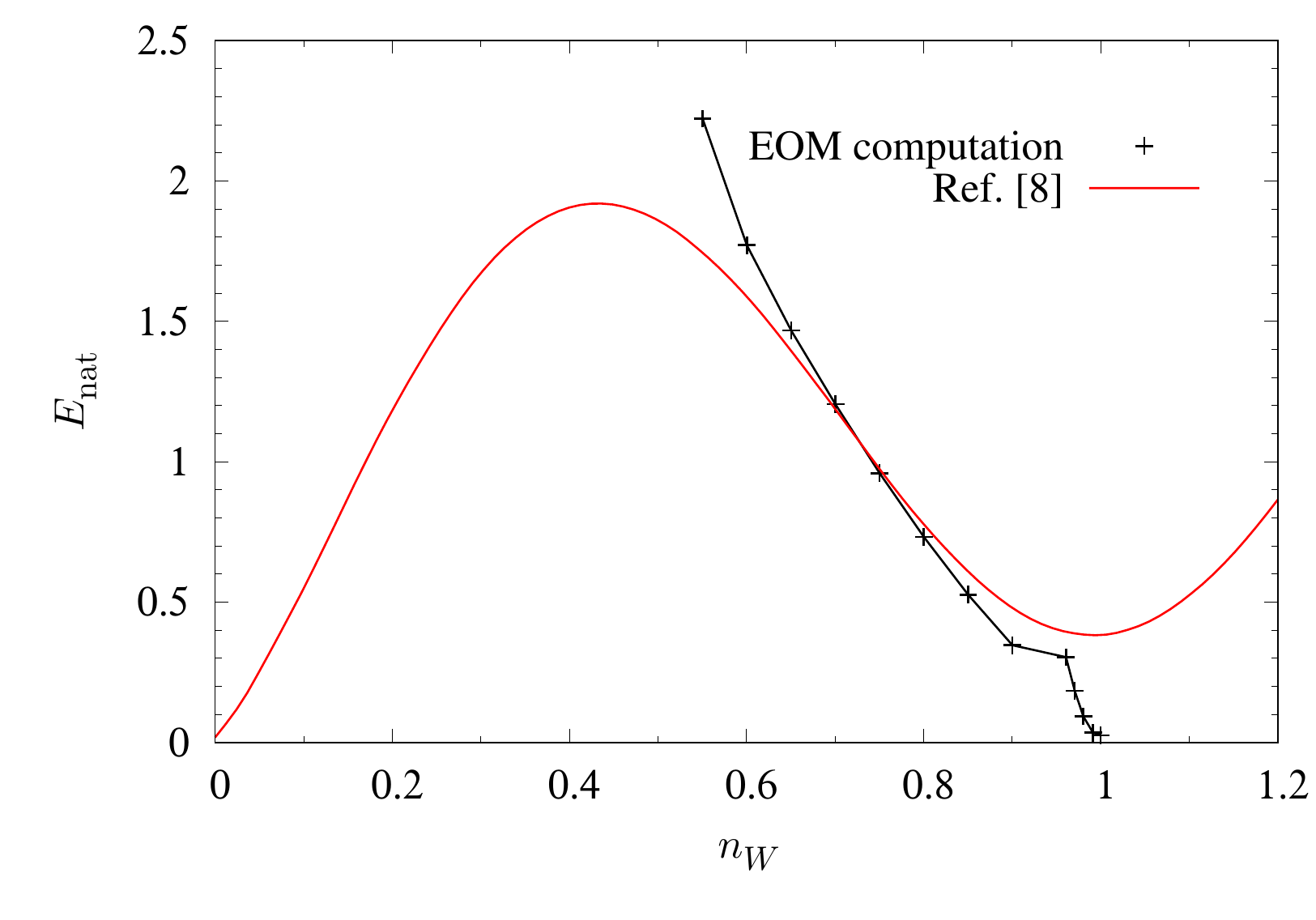}
    \caption{Dimensionless energy of EWS as a function of the
      modified Chern-Simons number ($n_W$) for $\Lambda=200\ \GeV$.
      The black dots represent solutions calculated with the EOM for
      the constrained problem, whereas the red curve is taken from
      Ref.~\cite{Criado:2020zwu}. }
    \label{fig:EB200}
  \end{center}
\end{figure}
In order to make an estimate, we attempt to flow the solution in field
configuration space from $(n_W,n_H)\simeq(1,0)$ to
$(n_W,n_H)\simeq(0,0)$, since this is the simplest and possible
by utilizing the additional Lagrange multiplier term \eqref{eq:Elambda}
and changing the Chern-Simons number from unity to zero.
This flow of solutions will estimate the height of the barrier of one
of the decay modes, which we estimate to be around 2 $\TeV$, see
Fig.~\ref{fig:EB200}.
We note that our flow solution became difficult to follow for
$n_W\lesssim0.55$; since the region in field configuration
space $0.6<n_W<0.9$ our results are quite similar to those of
Ref.~\cite{Criado:2020zwu}, their barrier height is most likely
correct.
Since $\sim2\ \TeV\gg 0.117\ \TeV$, with the latter being our result
for the EWS mass at $\Lambda=200\ \GeV$, we can utilize the instanton
approximation for the bound for estimating the lifetime of the EWS
\cite{Rubakov:1985it}:
\beq
\tau\approx M_W^{-1}e^{\frac{16\pi^2}{g^2}}
\approx 3.2\times 10^{134}\ {\rm s}\ .
\eeq
Notice that the approximate agreement between our results and those of
Ref.~\cite{Criado:2020zwu} stops to hold for $0.95<n_W\leq1$,
where the solution decreases significantly in energy.
For a more accurate estimate of the lifetime of the EWS, it is
necessary to find a bounce solution that can describe the unwinding of
both the Higgs field and the gauge fields. We propose the
spherically symmetric Ansatz 
\begin{equation}
    \begin{split}
    W_i^a(\textbf{x},\tau)&=\varepsilon_{aij}\hat{x}_j\frac{f_1(r,\tau)}{r}+(\delta_{ia}-\hat{x}_i\hat{x}_a)\frac{f_2(r,\tau)}{r}+\hat{x}_i\hat{x}_a\frac{b(r,\tau)}{r} \\
    W^a_0(\textbf{x},\tau)&=\hat{x}_af_0(r,\tau)\ , \qquad
    \Phi(x)=\frac{v}{\sqrt{2}}\left[\sigma(r,\tau)\mathds{1}_2+\tau^a\beta_a(r,\tau)\right]\ ,
    \end{split}
\end{equation}
where $\tau$ in the profile functions is Euclidean time. The Euclidean
equations of motion must be solved with the usual boundary
conditions of a bounce. Notice that the parametrization of the Higgs field is
nonsingular, which allows this solution to describe the two decay
channels mentioned above.

We now make an estimate of the height of the sphaleron barrier with a
different method. We note that if the Higgs field vanishes, the gauge
field configuration -- if pure gauge -- is both scale invariant and
deformable to a configuration with vanishing winding number. While
a scale transformation, i.e.~$r\to\mu r$, does not change the winding
number, a deformation $f_1\to\mu f_1$ with $f_2=\sqrt{f_1(1-f_1)}$
(from Eq.~\eqref{eq:enat}) and $b$ given by Eq.~\eqref{eq:bsol} with
$\sigma=0$ can change the winding number to zero. We assume here that the
gauge field configuration that we have found is so close to pure
gauge, that we can assume that the latter deformation 
holds, and we only need to disentangle the tails from the Higgs field
to be able to unwind the Skyrmion. We push the Higgs field to be zero
for $r\in[0,0.1]$ for the $\Lambda=200$ GeV configuration, and determine
the energy of this deformation
\begin{equation}
E_{\rm nat}^{\rm Higgs,\;deform} = \int\d r
\left[\frac{r^2}{2}(\sigma')^2+\beta r^2(\sigma^2-1)^2\right], \qquad
\sigma(r_{\rm min})=\sigma'(r_{\rm min})=0, \quad r_{\rm min}=0.1\ .
\end{equation}
We find $E_{\rm nat}^{\rm Higgs,\;deform}\sim 0.1491$, which
corresponds to $698$ GeV in physical units. This value is somewhat smaller than the
estimate made with the equations of motion shown in
Fig.~\ref{fig:EB200}, but we are here neglecting the fact that the
gauge fields are not exactly pure gauge. We thus believe that the
estimate $\approx 2$ TeV is quite reasonable.

\section{Quantization of collective coordinates}
\label{sec:ewcollcoor}

The collective coordinates of the EWSs can be
quantized following the procedure outlined in
Refs.~\cite{Adkins:1983hy, Weigel:2008zz} step-by-step, and using the
property \eqref{eq:rotations}.

\subsection{Rotations and isorotations}
The rotational collective coordinates of the EWS correspond to
the symmetries broken by the soliton and unbroken by the vacuum. 
This symmetry group consists of the spatial rotations and of the
diagonal isospin. These transformations act on $W_i$ according to
Eq.~\eqref{eq:rotations}, and leave $\Phi$ invariant.
If $A,L\in SU(2)$ are the matrices that parametrize rotations and diagonal
isorotations respectively, the Lagrangian for the corresponding
collective degrees of freedom is 
\begin{equation}
  L_{\text{rot}}=-E_0+\lambda\ \mathrm{tr}\big[\dot{(LA)}^{\dagger}\dot{(LA)}\big]\ ,\qquad
  \lambda=\frac{8\pi v^3}{3m_W^2\Lambda^2}\int_0^{\infty}\d r\ \left(2f_1^2+2f_2^2+b^2\right)\ ,
  \label{eq:moi}
\end{equation}
where $E_0$ is the classical energy of the EWS and
$\lambda$ is the moment of inertia and the coordinates and fields have
been rescaled as in Sec.~\ref{sec:numresults}.

\begin{figure}[!ht]
  \centering
  \includegraphics[width=0.49\textwidth]{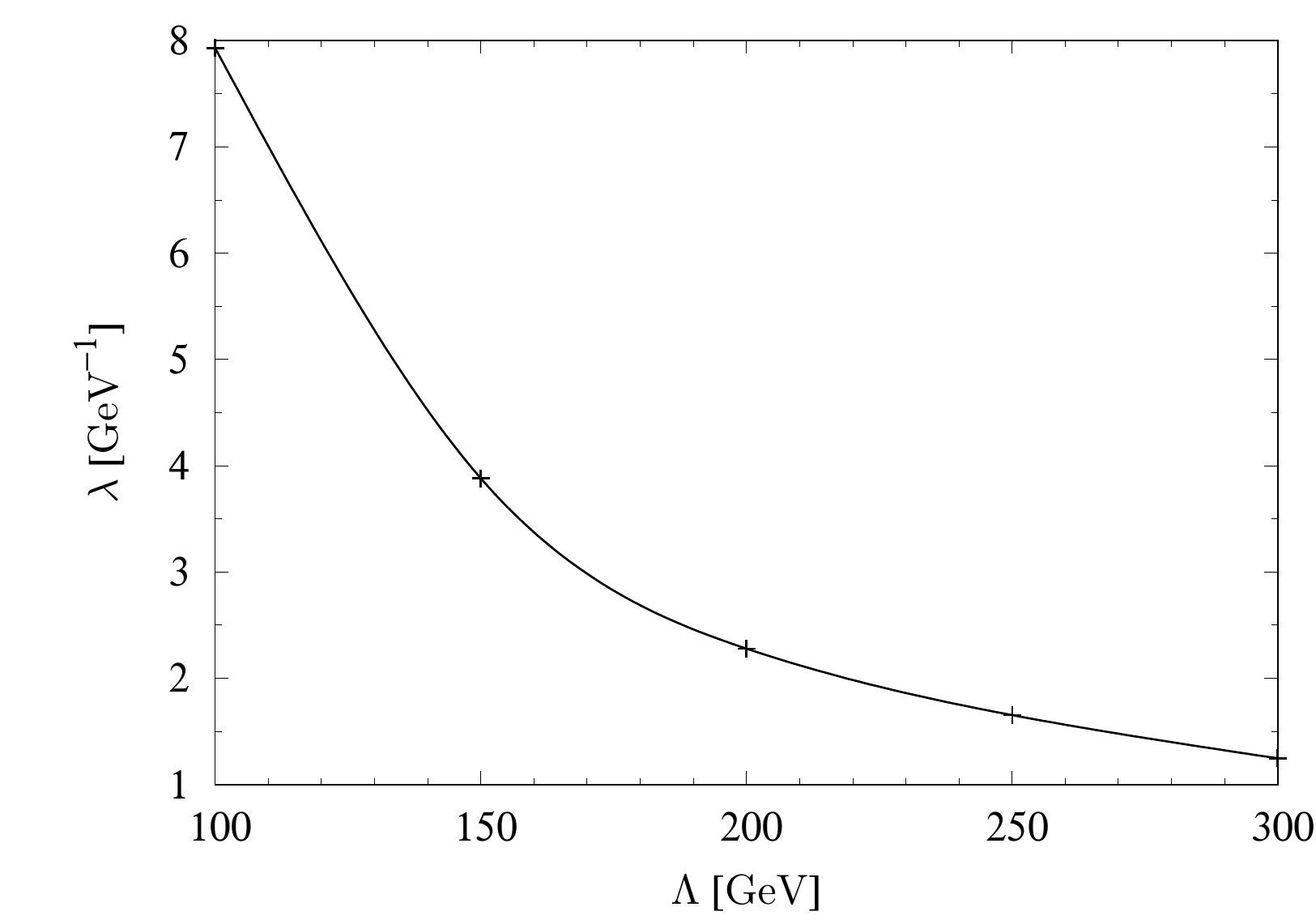}
  \caption{Moment of inertia, $\lambda$, of Eq.~\eqref{eq:moi} for the
  solutions of Fig.~\ref{fig:sols}. }
  \label{fig:moi}
\end{figure}
In Fig.~\ref{fig:moi} we show the moment of inertia of the EWS for the solutions of Fig.~\ref{fig:sols}. Like the size of the EWS, it shrinks as $\Lambda$ is increased. An approximate relation for moment of inertia as a function of $v$ and $\Lambda$ is given by
\begin{align} 
\lambda&\simeq\left(26.88 + \frac{\Lambda}{14.08\GeV}\right)
\frac{v^3}{m_W^2\Lambda^2}\ .
\label{fitsmrl_moi}
\end{align}
This Lagrangian describes a $(0+1)$-dimensional gauge theory with
local symmetry 
$A\mapsto U^{\dagger}A$, $L\mapsto LU$,
where $U\in SU(2)$ is time dependent.
We can use this gauge freedom to impose the condition $L=\mathds{1}_2$.
In this way, our Lagrangian further simplifies to
\begin{equation}
    \label{eq:rotlagrangian}
    L_{\text{rot}}=-E_0+\lambda\, \mathrm{tr}\big(\dot{A}^{\dagger}\dot{A}\big)\ .
\end{equation}
The quantization of a scalar field on a 3-sphere is well known
\cite{Adkins:1983ya,Weigel:2008zz} for the spherically symmetric
problem of the single Skyrmion. The generators of rotation and
weak isorotations can be constructed as
\begin{equation}
\label{eq:generators}
I_a=\i\lambda\, \mathrm{tr}\big(\dot{A}A^{\dagger}\tau_a\big), \qquad
J_a=-\i\lambda\, \mathrm{tr}\big(A^{\dagger}\dot{A}\tau_a\big)\ ,
\end{equation}
which are $\mathfrak{su}(2)$ algebra-valued, mutually commute and
identically satisfy $I^2=J^2$. 
Canonical quantization yields the rotational Hamiltonian
\begin{equation}
  H_{\mathrm{rot}}=E_0+\frac{I^2}{2\lambda}
  = E_0 + \frac{I(I+1)}{2\lambda}\ ,
  \label{eq:Hrot}
\end{equation}
where the eigenvalue of the squared operator of weak isospin is
$I(I+1)$.

\subsection{Breather mode}

Although technically it is not a collective coordinate, we also quantize the breather mode. We define a scale transformation as follows
\begin{equation}
\label{eq:scaling}
  W_i(\textbf{x})\longmapsto \eta W_i(\eta\textbf{x})\ , \qquad \Phi(\textbf{x})\longmapsto \Phi(\eta\textbf{x})\ ,
\end{equation}
where $\eta$ is a time-dependent quantity that parametrizes the transformation. Note that, according to this transformation law, spatial components of covariant derivatives scale as
\begin{equation}
  D_i\Phi(x)\longmapsto \eta D_i\Phi(\eta x) \ .
\end{equation}
In order to have $D_0\Phi(x)\longmapsto \eta D_0\Phi(\eta x)$, we must also let $W_0$ acquire a spherically symmetric nontrivial value, that we parametrize as
\begin{equation}
\label{eq:timecomponent}
  W_0(\textbf{x})=g\frac{\dot{\eta}}{\eta}\xi(\eta r)\ \hat{x}_a\frac{\tau_a}{2}\ .
\end{equation}
We now plug the transformed fields into the action
\eqref{eq:ewextended}, which yields a kinetic part due to
$\eta=\eta(t)$ and the potential part of the action pics up both
positive and negative powers of $\eta$.

First we need to determine the new profile function $\xi$, which turns
out to be exactly solved by $\xi=b$, with $b$ given by the solution
\eqref{eq:bsol} in terms of $f_1$, $f_2$ and $\sigma$.
Next we can determine the characteristic frequency of oscillations,
which is the breather mode.
Expanding in small perturbations $\epsilon=\eta-1$, we obtain the
breather frequency
\beq
\omega_{c}=\sqrt{\frac{2B_{\text{nat}}+12C_{\text{nat}}}{D_{1,\rm nat}+D_{2,\rm nat}}}\ ,
\eeq
with the functionals
\begin{align}
  B_{\rm nat} &= \int_0^\infty\d r
  \left[\frac{r^2}{2}(\sigma')^2 + \sigma^2\left(f_1^2+f_2^2+\frac{b^2}{2}\right)\right],\non
  C_{\rm nat} &= \int_0^\infty\d r\;\beta r^2(\sigma^2-1)^2,\non
  D_{1,\rm nat} &= \int_0^\infty r^2\d r\bigg\{
  2\alpha\bigg[\left(f_1' - \frac{2b}{r}f_2\right)^2 + \left(f_2' + (2f_1-1)\frac{b}{r}\right)^2\bigg]\non
  &\phantom{=\int_0^\infty r^2\d r\bigg\{\ }
  +2(f_1^2+f_2^2)\sigma^2\left((\sigma')^2 + \frac{\sigma^2}{r^2}b^2\right)\bigg\},\non
    D_{2,\rm nat} &= \int_0^\infty r^2\d r\left(r^2(\sigma')^2 + b^2\sigma^2\right).
\end{align}
For further details on this derivation, see App.~\ref{app:breather}.

\begin{figure}[!ht]
  \centering
  \includegraphics[width=0.6\textwidth]{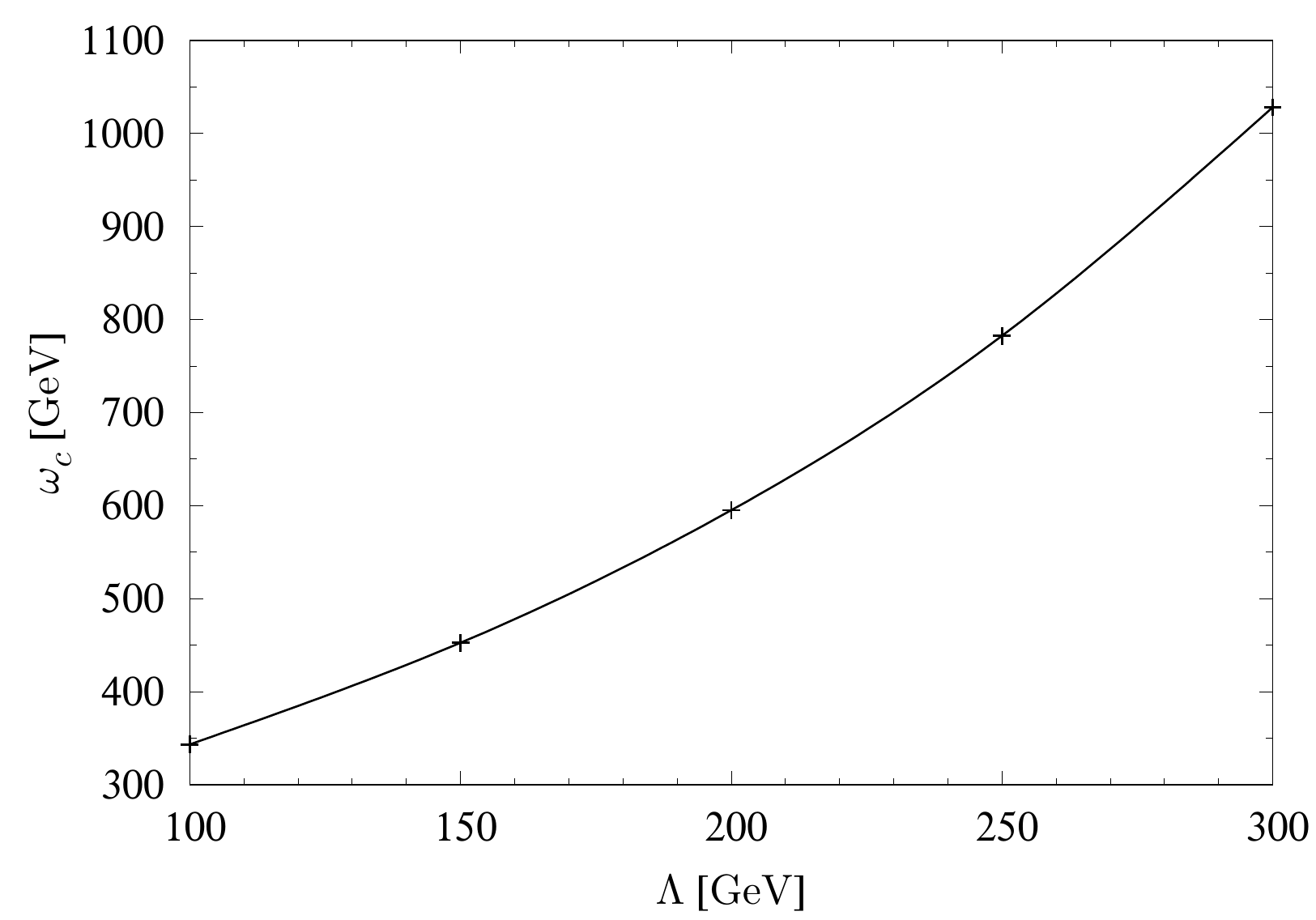}
  \caption{The breather frequency, $\omega_c$ (in $\GeV$) for the
    solutions of Fig.~\ref{fig:sols}. }
  \label{fig:omegac}
\end{figure}
Fig.~\ref{fig:omegac} shows the breather frequency for the solutions
of Fig.~\ref{fig:sols} in $\GeV$, which in physical units grows nearly
linearly with $\Lambda$.
A fit for $\omega_c$ is:
\beq
\omega_c \simeq 277.917\, \GeV - 0.23092 \Lambda + 0.00907898 \frac{\Lambda^2}{\GeV}\ .
\eeq

\subsection{Parity}
\label{sec:parity}
Let us now consider the action of parity on the collective modes.
For a time-independent vector field, parity changes the overall sign
of both the argument and of its spatial part (in 3-dimensional space),
and the argument of the temporal part 
\begin{equation}
  P:\ W_0(\textbf{x})\longmapsto W_0(-\textbf{x})\ , \qquad
  W_i(\textbf{x})\longmapsto -W_i(-\textbf{x})\ .
\end{equation}
Performing this transformation on the Skyrmion field, one sees that
its effect consists in changing the sign of $f_2$ and $b$.
It is easy to see that this transformation cannot be reabsorbed by a
transformation of any of the collective coordinates $A$ and
$\varepsilon$, and that the EWS defined above does not seem to have a
definite parity.
The action of parity can be interpreted as follows: the Skyrmion is
not sent to a rotated version of itself, but rather to a new field
configuration that we can call the \textit{anti-Skyrmion}.
We can add a further quantum number $\uparrow$\ $\downarrow$ that
specifies whether the field configuration is a Skyrmion or an
anti-Skyrmion, and that transforms as 
\begin{equation}
  P\ket{\uparrow}=\ket{\downarrow}\ , \qquad
  P\ket{\downarrow}=\ket{\uparrow}\ .
\end{equation}
The parity operator $P$ is diagonal in the rotated basis
\begin{equation}
\ket{\pm}=\frac{\ket{\uparrow}\pm\ket{\downarrow}}{\sqrt{2}}\ .
\end{equation}
The parity operator $P$ commutes with all the rotation generators
\eqref{eq:generators}. The profile functions $f_1(r)$, $f_2(r)$ and
$b(r)$ must be \textit{formally} promoted to commuting operator-valued
functions $\hat{f}_1(r)$, $\hat{f}_2(r)$ and $\hat{b}(r)$ such that 
\begin{equation}
\label{eq:parity}
\begin{split}
  &\hat{f}_1\ket{\uparrow}=+f_1\ket{\uparrow}\ , \qquad
  \hat{f}_2\ket{\uparrow}= +f_2\ket{\uparrow}\ , \qquad
  \hat{b}\ket{\uparrow}=+b\ket{\uparrow}\ , \\
  &\hat{f}_1\ket{\downarrow}=+f_1\ket{\downarrow}\ , \qquad
  \hat{f}_2\ket{\downarrow}=- f_2\ket{\downarrow}\ , \qquad
  \hat{b}\ket{\downarrow}=-b\ket{\downarrow}\ ,
\end{split}
\end{equation}
and
\begin{equation}
  P^{\dagger}\hat{f}_1P=+\hat{f}_1\ , \qquad
  P^{\dagger}\hat{f}_2P=-\hat{f}_2\ , \qquad
  P^{\dagger}\hat{b}P=-\hat{b}\ .
\end{equation}
In general, any observable $O$ will be a function of $A$,
$\varepsilon$, and the profile functions $f_1, f_2, b$.
From Eq.~\eqref{eq:parity} we have
\begin{equation}
\begin{split}
    O(A, \varepsilon; \hat{f}_1, \hat{f}_2, \hat{b})\ket{\uparrow}&=O(A, \varepsilon; f_1, f_2, b)\ket{\uparrow}\equiv O^{\uparrow}(A, \varepsilon)\ket{\uparrow}\ ,\\
    O(A, \varepsilon; \hat{f}_1, \hat{f}_2, \hat{b})\ket{\downarrow}&=O(A, \varepsilon; f_1, -f_2, -b)\ket{\downarrow}\equiv O^{\downarrow}(A, \varepsilon)\ket{\downarrow}\ ,
\end{split}
\end{equation}
where all the other quantum numbers have been omitted. We define the parity operator so that it leaves $A$ and $\varepsilon$ invariant
\begin{equation}
  P^{\dagger}AP=A\ , \qquad
  P^{\dagger}\varepsilon P=\varepsilon\ .
\end{equation}
The action of $P$ on $A$ has been chosen such that the angular
momentum is a pseudovector and the weak isospin operator is diagonal
and parity-even.
Note that the moment of inertia, $\lambda$, and the characteristic
frequency of the breather, $\omega_c$, are both parity-invariant
quantities.

\subsection{Spin and statistics}
To determine whether the EWS should be quantized as a boson or as a
fermion, we need to study the action of the rotation group on the
corresponding field configuration.
Following Manton \cite{Manton:1983nd}, we impose the radial gauge
condition $W_r=0$; i.e.~we need to find a gauge transformation that
satisfy
\begin{equation}
UW_rU^{\dagger}-\i\partial_rUU^{\dagger}=0\ .
\end{equation}
Using $W_r(\textbf{x})=\hat{x}_i W_i(\textbf{x})=g\frac{\tau_a}{2}\hat{x}_aF(r) $, we find that the unitary matrix $U$ must satisfy the parallel transport equation
\begin{equation}
\partial_rU=-\i UW_r \quad \implies \quad
U(\textbf{x})=\mathrm{exp}\left(-\i g\frac{\tau_a}{2}\hat{x}_a\int_0^{|x|}\d r'\ F(r')\right)\ .
\end{equation}
The lower integration limit in the exponent of $U(\textbf{x})$ has
been chosen in order to have a function that is single-valued at the
origin.
Now the gauge is completely fixed up to a global gauge rotation.
The Higgs field at infinity takes the form
\begin{equation}
\Phi_{\infty}(\hat{x})=\mathrm{exp}\left(-\i g\frac{\tau_a}{2}\hat{x}_a\int_0^{\infty}\d r\ F(r)\right)\ .
\end{equation}
The asymptotic behavior in Eq.~\eqref{eq:asymptotic} ensures
finiteness of the exponent.
Introducing the coordinates $\theta\in[0,\pi]$ and $\varphi\in [0,2\pi]$ on $S^2$, we can write
\begin{equation}
    \tau_a\hat{x}_a(\theta, \varphi)=
    \begin{pmatrix}
    \cos\theta  &  \sin\theta\ e^{-\i\varphi} \\
    \sin\theta\ e^{+\i\varphi} & \cos\theta
    \end{pmatrix} \ .
\end{equation}
A $2\pi$ rotation defines a closed loop in the space of continuous functions $S^2\longrightarrow S^3$, parametrized by
\begin{equation}
  \Phi_{\infty}(\hat{x},\sigma)=A(\sigma)\Phi_{\infty}(\hat{x})A^{\dagger}(\sigma)\ , \qquad
  A(\sigma_i)=A(\sigma_f)=\mathds{1}_{2}\ .
\end{equation}
We need to determine whether or not this loop is contractible. As the
loop is closed, we can compactify the $\sigma$ coordinate, and see the
loop as an element of the space of continuous functions $S^3\to S^3$,
of which we need to know the homotopy class. This can be computed as
in Ref.~\cite{Manton:2004tk}. In order to simplify our calculation, we
choose $A(\sigma)$ so that it describes a circular loop around the
$\hat{x}_3$ axis. In this way, $\Phi_{\infty}(\hat{x},\sigma)$ can be
obtained simply by the replacement
\begin{equation}
    \Phi_{\infty}(\hat{x}(\theta,\varphi))\longrightarrow\Phi_{\infty}(\hat{x}(\theta,\varphi+\sigma))\ ,
\end{equation}
where $\sigma$ has been rescaled so that $\sigma\in[0,2\pi]$. The homotopy class of the loop can be computed from the integral
\begin{equation}
    \frac{1}{24\pi^2}\int\d\mu(\theta,\varphi,\sigma)\mathrm{tr}\left(\d\Phi_{\infty} \Phi_{\infty}^{-1}\wedge \d\Phi_{\infty} \Phi_{\infty}^{-1}\wedge \d\Phi_{\infty} \Phi_{\infty}^{-1}\right)\ ,
\end{equation}
where $\d\mu(\theta,\varphi,\sigma)$ is a measure whose exact form is
not needed. We can immediately see that this integral is zero, as,
thanks to our choice of the loop, the derivatives with respect to
$\varphi$ and $\sigma$ are equal, and thus they cancel due to
antisymmetry. We conclude that our loop is contractible, and thus the
EWS must be quantized as a boson. 
This means that the spin and weak isospin eigenvalues $I$ of
Eq.~\eqref{eq:Hrot} must be integers.

\subsection{Discussion}

\begin{figure}[!ht]
  \centering
  \includegraphics[width=0.6\textwidth]{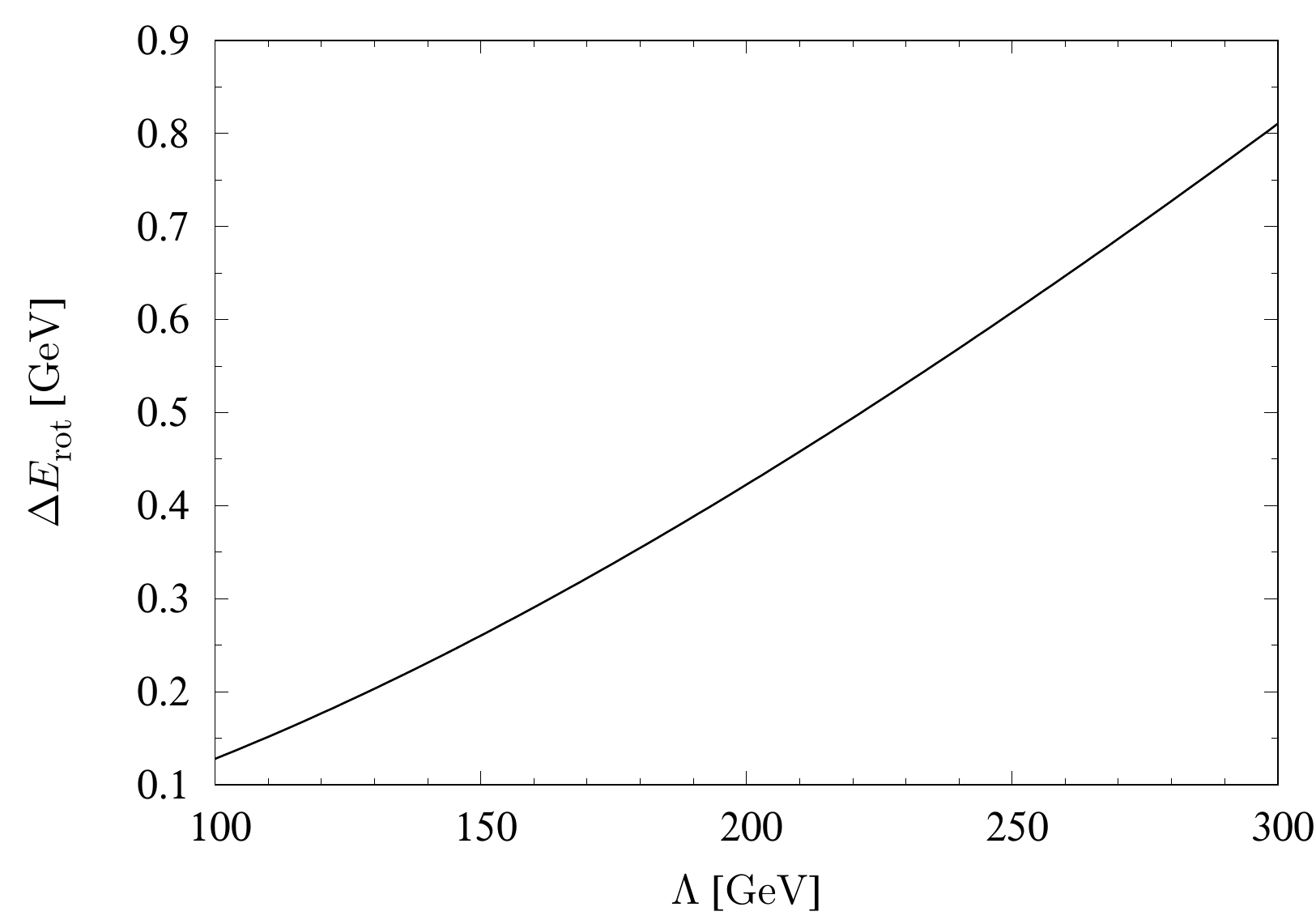}
  \caption{The rotational energy gap, $\Delta E_{\rm rot}$, in GeV,
    for the solutions of Fig.~\ref{fig:sols}. }
  \label{fig:rotgap}
\end{figure}

From the previous subsections, it follows that the Hilbert space of a
static EWS is spanned by states of the form 
\begin{equation}
    \ket{I=J,I_3,J_3, n, \uparrow\downarrow}\ .
\end{equation}
The energy of such states is
\begin{equation}
    E_{I=J,n}=E_0+\frac{I(I+1)}{2\lambda}+\left(n+\frac{1}{2}\right)\omega_{c}\ .
\label{enegspec}
\end{equation}
The spin and weak isospin are always integer, and the states are
bosonic. We can now discuss the order of magnitude of the energy gaps
for the physical parameters considered in solutions of
Fig.~\ref{fig:sols}; i.e.~$100\leq \Lambda \leq 300$.
We consider the ground state $E_{0,0}$, the first rotationally excited
state $E_{1,0}$ and the first vibrationally excited state $E_{0,1}$ in
Eq.~\eqref{enegspec}. For $E_0$, $\lambda$ and $\omega_c$ we use the
numerical fits obtained earlier. The rotational energy gap which we
define as 
\beq
\Delta E_{\rm rot} = E_{1,0}-E_{0,0} = \frac{1}{\lambda}\ ,
\eeq
is plotted in Figure \ref{fig:rotgap}.
 
Thus the rigid rotor approximation is valid, as the rotational energy
gap is two orders of magnitude smaller than the mass of the EWS.
For thermal energies much smaller than $\Delta E_{\rm rot}\sim$
GeV the soliton is in its  ground state.  
The vibrational energy gap is defined as
\beq
\Delta E_{\rm vib} = E_{0,1}-E_{0,0} = \omega_c\ .
\eeq
The EWS with $\Lambda\in[100,300]\ \GeV$ thus possesses the properties
\beq
\Delta E_{\rm rot} \ll E_{0,0} \sim \Delta E_{\rm vib} \ .
\eeq
For practical purposes, this yields important implications.
First, the rigid rotor approximation is valid, as the rotational
energy gap is two orders of magnitudes smaller than the mass of the
EWS.
The vibrational energy gap is of order or even bigger than the EWS
mass, so we do not have to consider these effects as long as the
soliton approximation is valid.
For thermal energies much smaller than $\Delta E_{\rm rot}\sim$ GeV the
soliton is in its spin-zero ground state.

\section{Interaction with leptons}\label{sec:eqleptons}

\subsection{Lepton number}
The fractional part of the lepton number of a soliton can be obtained through the Goldstone-Wilczek adiabatic method \cite{Goldstone:1974gf, Aitchison:1985pp}, which yields the expression
\begin{equation}
\label{eq:jlepton}
\langle{j^{\mu}}_{\mathrm{l.o.}}\rangle=\frac{1}{24\pi^2}\varepsilon^{\mu\nu\rho\sigma}\mathrm{tr}\left(H^{\dagger}D_{\nu}HH^{\dagger}D_{\rho}HH^{\dagger}D_{\sigma}H+\frac{3\i}{2}H^{\dagger}W_{\nu\rho}D_{\sigma}H\right).
\end{equation}
This is exactly the anomalous current whose charge is appears in
\eqref{eq:topinv}. As explained in Sec.~\ref{sec:modelews}, this
quantity does not exist for field configurations for which
$s(\textbf{x})=0$ at one (or more) point(s), because the derivative
expansion of the fermion current needed to obtain this result, works
under the assumption of small field gradients. This condition
is violated when the fermion mass term goes to zero at one point,
leading to a pathological behavior in the results of the calculation.

For field configurations with $H=\mathds{1}_2$, the lepton charge
computed from \eqref{eq:jlepton} is equal to its Chern-Simons
number. Substituting the profile functions of the Skyrmion field
configuration as expressed in equation \eqref{eq:profile} into the
Chern-Simons number, one sees that the fermion number exists, and its
value is 
\begin{equation}
  \hat{n}_{\ell}=\hat{n}_{\mathrm{CS}}=
  -\frac{g^3}{2\pi}\int\frac{1}{r}\d r\,\big(\hat{f}_1^2+\hat{f}_2^2\big)\hat{b}
  +\frac{g^2}{2\pi}\int\frac1r\d r\,\big(r\hat{f}_2\hat{f}_1'-r\hat{f}_1\hat{f}_2'+2\hat{f}_1\hat{b}\big)\ .
\end{equation}
where we used the notation introduced in Sec.~\ref{sec:parity}.
This quantity is not an ordinary number, but an operator that acts on
the Hilbert space of collective coordinates. The only quantum number
that is relevant to the value of $n_{\ell}$ is parity. If $n_{\rm CS}$ is
the Chern-Simons number calculated on the soliton state, we obtain
that, in the basis $\ket{\uparrow\downarrow}$, the lepton number is a
matrix
\begin{equation}
  (\hat{n}_{\ell})_{\uparrow\downarrow}=\begin{pmatrix}
  n_{\rm CS} & 0 \\
  0 & -n_{\rm CS}
  \end{pmatrix}.
\end{equation}
Note that parity and lepton number cannot be diagonalized simultaneously.

\subsection{Spectral flow}
As the Chern-Simons number of the EWS is, in general, a non-integer
quantity, it cannot be equal to the number of leptons that is emitted
(or absorbed) during the destruction (or production) of an EWS. It has
been shown that this latter quantity is equal to the net number of
fermion levels of the Dirac Hamiltonian 
\begin{equation}
  \label{eq:skyrmdirac}
  H_{\rm Dirac}=\i\gamma^0\gamma^i(\partial_i-\i W_iP_L)-\gamma^0y(\Phi P_R+\Phi^{\dagger}P_L)\ ,
\end{equation}
that cross zero energy during an adiabatic interpolation that connects
the initial and final states \cite{Christ:1979zm}. As the quantity we
are looking for is a gauge-invariant integer, we reasonably conjecture
that the number of leptons produced in such a process is a function
$f(\Delta n_H)$, where $n_H$ has been defined in Eq.~\eqref{eq:winding}
and is the only integer and gauge-invariant quantity characterizing
the interpolation. The other potential quantity that may play this
role is the integer part of $\Delta n_{\rm CS}$, but by a suitable
gauge transformations can always be turned into $\Delta n_H$. This
function must satisfy the conditions 
\begin{equation}
  f(n)\in\mathbb{Z}\ , \qquad f(0)=0\ , \qquad f(kn)=kf(n)\ , \qquad
  \forall k,n\in \mathbb{Z}\ .
\end{equation}
From this conditions it follows that
\begin{equation}
	f(n)=a\cdot n \qquad \mathrm{for\ some} \qquad a\in\mathbb{Z}
\end{equation}
This conjecture is known to be true in the case in which the initial
weak field is pure gauge with nonzero winding number and the Higgs
scalar is frozen at its vacuum expectation value
\cite{Christ:1979zm}. Moreover, in Ref.~\cite{Farhi:1995aq} it is
shown that this conjecture is true also in the case of the Skyrmion
and of the EWS with non-dynamical Higgs. In these cases, the constant
$a$ is equal to the number of leptons that satisfy $m_{\ell}R\ll 1$,
where $R$ is the Skyrmion radius. 

When the Higgs is not frozen, the previous results can be taken as a
point of departure for a new argument. Using the same approach of
Ref.~\cite{Farhi:1995aq}, we have to study the spectral flow under a
suitable interpolation that connects the EWS with dynamical Higgs to
the EWS in the limit $m_h\to\infty$.
Unfortunately, this interpolation has a peculiarity
that may invalidate the previous results. In order to see what happens
during the interpolation, let us compare the squares of the Dirac
Hamiltonian in the case of frozen, and of dynamical Higgs 
\begin{subequations}
  \begin{align}
    \label{eq:frozen}
    H^2&=-D^2+m_{\ell}^2-\frac{1}{2}W_{ij}\sigma^{ij}P_L\ ,
    \qquad\qquad\quad\ \, \mathrm{(frozen)}\\
    \label{eq:dynamical}
    H^2&=-D^2+y^2s^2+\i y\slashed{\partial}s-\frac{1}{2}W_{ij}\sigma^{ij}P_L\ . \qquad
    \mathrm{(dynamical)}
  \end{align}
\end{subequations}
In both cases, the field strength term is of order $1/R^2$, and must
be compared with the other terms to see if a zero energy bound state
may, at least in principle, appear during the interpolation. While in
Eq.~\eqref{eq:frozen} the free Hamiltonian has a mass gap
$m_{\ell}$ and the gauge fields can be considered as a perturbation
around it, in Eq.~\eqref{eq:dynamical} any possible mass gap
depends on the shape of $s(r)$, and can be arbitrarily close to
zero. This makes the study of the spectral flow of the Dirac Hamiltonian
particularly difficult.

\section{Long-distance interaction}
\label{sec:longdistance}
In this section we consider the interaction between two distant rotating
EWSs, neglecting vibrational modes. We expect a
contribution to the semiclassical interaction energy for each
fundamental degree of freedom: the weak and hypercharge gauge bosons,
the Higgs scalar, and the leptons. In the end of this section, we will
see that the contribution of the hypercharge gauge fields is dominant
even at the classical level. Some general considerations on
long-distance interactions between solitons are given in
App.~\ref{app:intpot}. 

\subsection{Higgs and gauge fields}

Let us consider two spherically symmetric Skyrmion field
configurations with centers at $\textbf{x}_1$ and $\textbf{x}_2$,
characterized by the rotational collective coordinates $A,B\in
SU(2)$. The superposition of them has the form
\begin{equation}
\begin{split}
	W_i(\textbf{x}; A,B)&=AW_i(\textbf{x}-\textbf{x}_1)A^{\dagger}+BW_i(\textbf{x}-\textbf{x}_2)B^{\dagger}\ , \\ h(\textbf{x})&=h(|\textbf{x}-\textbf{x}_1|)+h(|\textbf{x}-\textbf{x}_2|) \ .\label{eq:asymp_ansatz}
\end{split}
\end{equation}
This field approximately satisfies the equations of motion when
$|\textbf{x}_{12}|\gg R$ with $R$ being the Skyrmion radius. Due to 
exponential suppression of the fields at spatial infinity, the leading
order contribution to the interaction potential is given by the
quadratic interaction terms in the static energy functional
$V_{\mathrm{int}}=V_{\mathrm{int}}^{(h)}+V_{\mathrm{int}}^{(W)}$ where
\begin{subequations}
    \label{eq:interaction}
    \begin{align}
        V_{\mathrm{int}}^{(h)}(\textbf{x}_{12})&=-\int\d^3\textbf{x}\ h^{(1)}(-\nabla^2+m_h^2)h^{(2)}\ ,\\
        V_{\mathrm{int}}^{(W)}(\textbf{x}_{12})&=-D_{ab}(B^{\dagger}A) \int\d^3\textbf{x}\ W_{ai}^{(1)}(-\nabla^2\delta_{ij}+\partial_i\partial_j+m_W^2\delta_{ij})W_{bj}^{(2)}\ .
    \end{align}
\end{subequations}
We used the same notation of App.~\ref{app:intpot} to denote the
asymptotic fields at large distances. To obtain the interaction
energy, we need the source terms for $h$ and $W$ (see
App.~\ref{app:asymptotic_behavior}):
\begin{subequations}
    \begin{align}
        S^{(h)}(\textbf{x})&=\frac{c_hv}{m_h}\delta^{(3)}(\textbf{x})\ ,\\
        S_{ai}^{(W)}(\textbf{x})&=-\frac{c_{\alpha}}{m_W^2}\varepsilon_{aij}\partial_j\delta^{(3)}(\textbf{x})\ .
    \end{align}
\end{subequations}
Substituting into Eq.~\eqref{eq:interaction} and enforcing the delta
constraint, one obtains the potentials 
\begin{subequations}
    \begin{align}
    \label{eq:vinth}
        V_{\mathrm{int}}^{(h)}(r)&=-\left(\frac{c_hv}{m_h}\right)^2\frac{e^{-m_hr}}{4\pi r}\ ,\\
    \label{eq:vintw}
        V_{\mathrm{int}}^{(W)}(\textbf{x})&=-D_{ab}(B^{\dagger}A) \left(\frac{c_{\alpha}}{m_W^2}\right)^2 (-\nabla^2 \delta_{ab}+\partial_a \partial_b) \frac{e^{-m_W r}}{4 \pi r}\ ,
    \end{align}
\end{subequations}
where $r=|\mathbf{x}|$ is the distance between the Skyrmions. We see that the
interactions mediated by the Higgs and gauge fields have a range $\sim
1/v$. The Higgs exchange interaction potential \eqref{eq:vinth} does
not depend on the orientation of the solitons, which means that, if we
see it as an operator acting on the two-soliton Hilbert space, its
effect is state-independent. Instead, the operator \eqref{eq:vintw}
depends on the relative orientation of the solitons, and its effect
depends on the value of the relative angular momentum
$J_{\mathrm{rel}}=J_1-J_2$. As a consequence, two Skyrmions in the
ground state feel Higgs exchange interaction but no vector boson
exchange interaction.

\begin{figure}[!ht]
  \centering
  \includegraphics[width=0.49\textwidth]{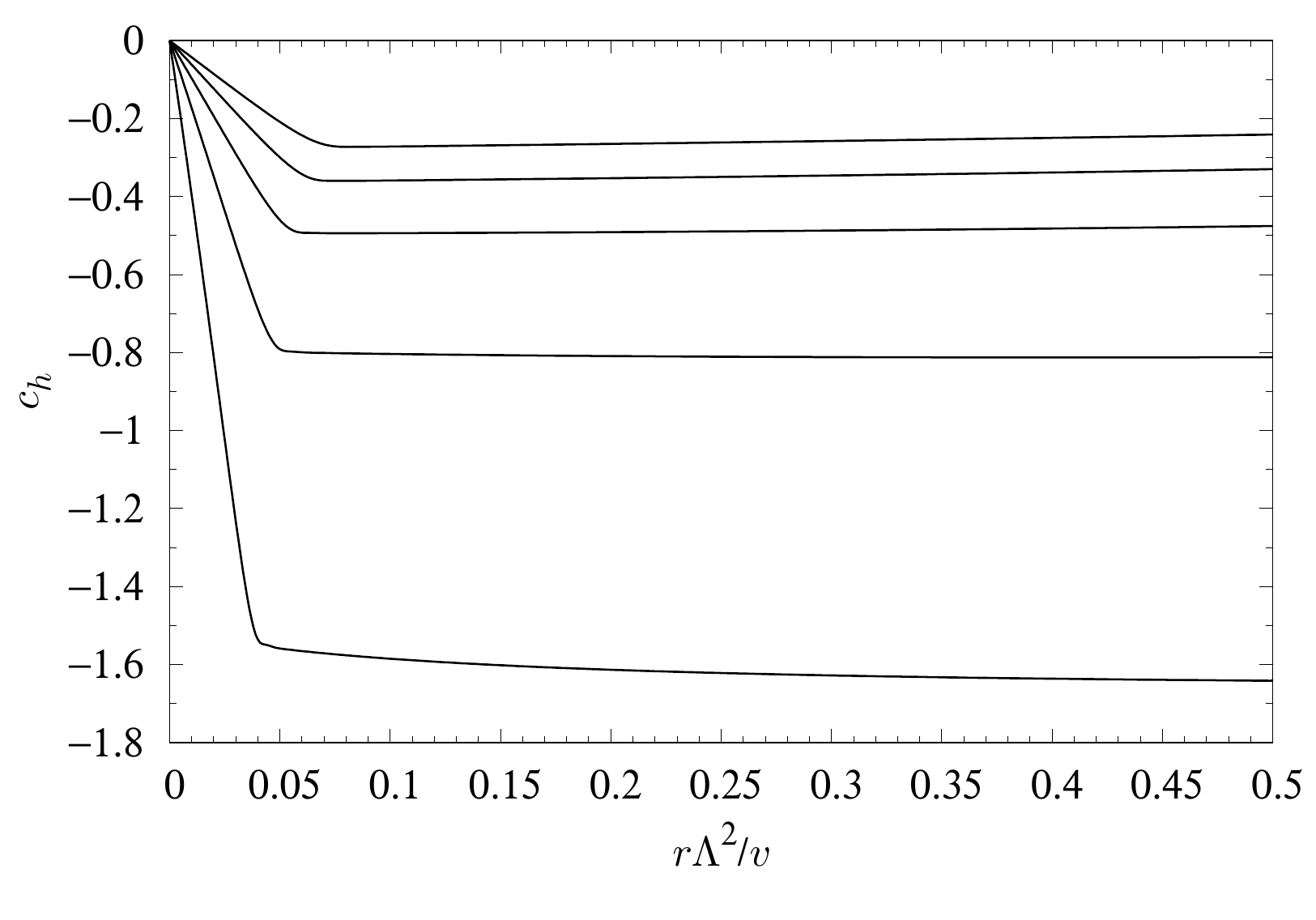}
  \includegraphics[width=0.49\textwidth]{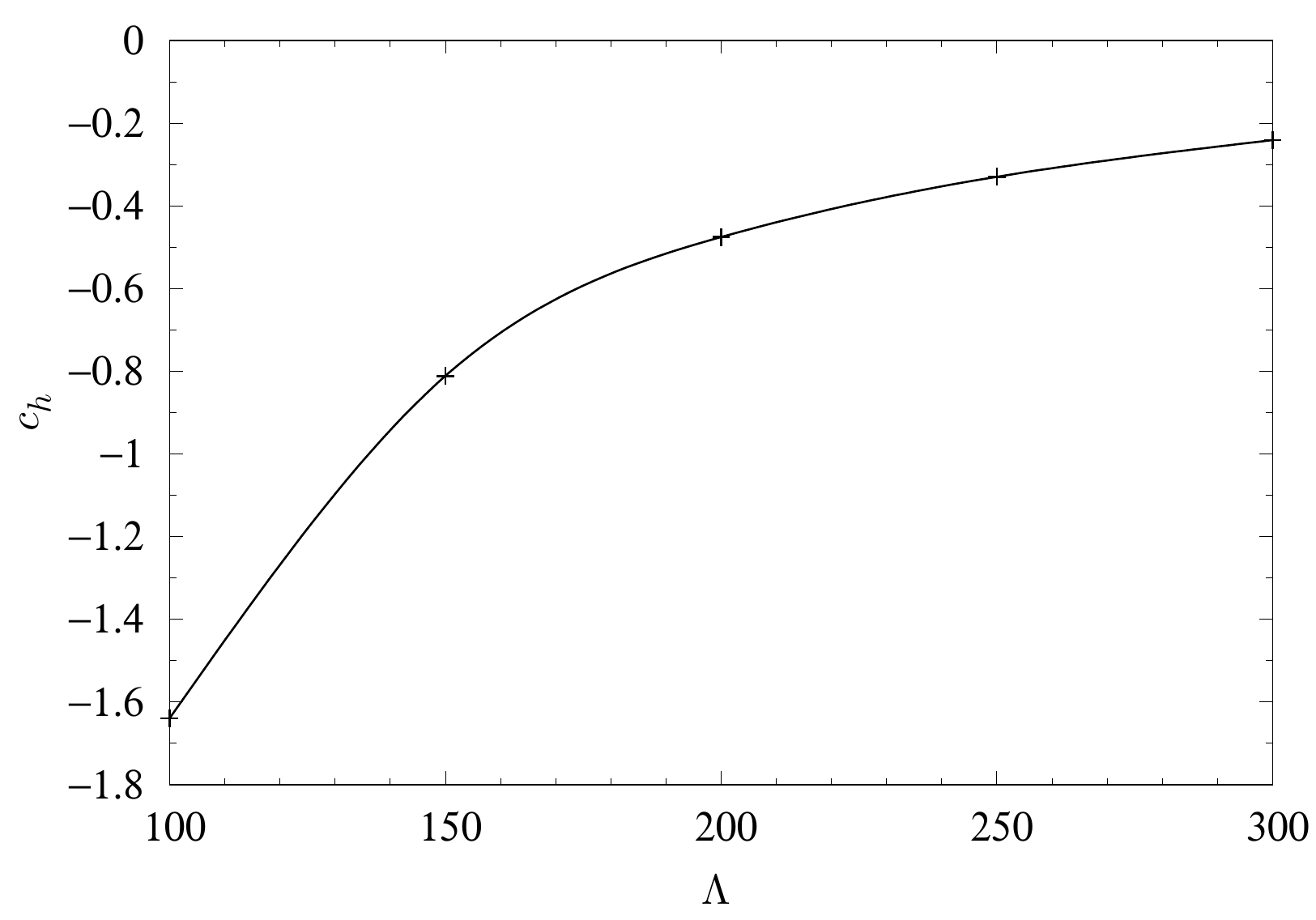}
  \caption{ The coefficient $c_h$ of Eq.~\eqref{eq:asymphiggs} computed with
    the numerical solutions presented in Fig.~\ref{fig:sols}.  }
  \label{fig:ch}
\end{figure}
In Fig.~\ref{fig:ch} we show the coefficient of the asymptotic
function, that the Higgs field tends to, see
Eq.~\eqref{eq:asymphiggs}.
In the left panel, the convergence of the solution to the asymptotics
is shown, whereas the extracted value is shown in the right panel.
Fitting the $\Lambda$ behavior of the coefficient $c_h$, we get
\beq
c_h \approx -\frac{28.64\,\GeV}{\Lambda}
-\left(\frac{116.5\,\GeV}{\Lambda}\right)^2\ .
\eeq

\subsection{Electromagnetic properties}
\label{em}
The temporal and spatial components of the electromagnetic current are
\begin{align}
  \label{eq:emcurrent}
  J^0_{\mathrm{em}}&=\varepsilon_{3ab}W_i^a\dot{W}_i^b \ , \\
  J^i_{\mathrm{em}}&=\varepsilon_{3ab}\left(W_j^a\partial_iW_j^b-W_j^a\partial_jW_i^b\right)+gW_i^3\left[(W_j^1)^2+(W_j^2)^2\right]+gW_j^3\left[W_j^1W_i^1+W_j^2W_2^1\right] \ .\nonumber
\end{align}
Using these expressions, we can compute the electric charge, and the
electric and magnetic dipole moments of a rigidly rotating
EWS. Recalling that, due to the symmetries, a rotation acts on the EWS
fields as in Eq.~\eqref{eq:rotations}, we obtain, for the electric charge 
\begin{equation}
  Q_{\rm em}=\varepsilon_{3ab}\int\d^3\textbf{x}\ W_i^a\dot{W}_i^b=\frac{8\pi\i}{3}\int_0^{\infty}\d r\left(2f_1^2+2f_2^2+b^2\right)\ \mathrm{tr}\big(\dot{A}A^{\dagger}\tau^3\big)=I_3\ .
\end{equation}
For the electric dipole moment and for the magnetic dipole moment, we can use the fact that they both behave as vectors under rotations: we can compute them for $A=\mathds{1}_{2}$, and rotate them afterwards. The electric dipole moment turns out to be identically zero
\begin{equation}
  d_{\mathrm{em}}^i=\varepsilon_{3ab}\int\d^3\textbf{x}\ x_iW_j^a\dot{W}_j^b=0\ .
\end{equation}
Instead, the magnetic dipole is
\begin{align}
    m_i&=\varepsilon_{ijk}\int\d^3\textbf{x}\ x_jJ_{\mathrm{em}}^k \non &=\frac{4\pi}{3}\delta_{i3}\int_0^{\infty}\d r\left\{[2f_1^2+b(b-f_2)]-\frac{1}{5}f_1[13(f_1^2+f_2^2)+7b^2]\right\}.
\end{align}
For a rotating EWS, the magnetic dipole moment is thus proportional to
$D_{i3}(A)$. If seen as an operator acting on the Hilbert space of the
collective coordinates, $D_{i3}(A)$ vanishes in the ground state due
to rotational invariance. Contrarily to the classical solution, the
quantized EWS does not possess a magnetic dipole moment. At the
classical level, this computation of the magnetic dipole moment is
equivalent to solving the equations of motion of the hypercharge field
$B_{\mu}$ for $g'\ll g$ in the EWS background through a multipole
expansion. Of course, this procedure neglects the backreaction of the
hypercharge field, and a more careful analysis should treat the
hypercharge and the weak gauge fields on equal footing.

\subsection{Leptons}
The interaction energy due to the leptons can be obtained by integrating
out the lepton fields and repeating the calculation of the previous
sections on the resulting quantum effective action. The path
integration of the Fermi variables has been performed in
Refs.~\cite{DHoker:1984izu, DHoker:1984mif}. The only effect of the
leptons on the quadratic part of the initial Lagrangian is to
renormalize the coupling constants. In order to see the contribution
of the lepton exchange to the interaction energy (that we expect to
have a range $\sim m_{\ell}^{-1}$) we need to take into account
higher-order terms that are suppressed by the distance between the
EWSs. We will leave such calculation for future work.

\section{EWS phenomenology}
\label{sec:phenomenology}

\subsection{Collider constraints on the Skyrme term}
We obtain some experimental constraints on the coefficient of the
Skyrme term from analysis of the data collected by the CMS
detector at the LHC (CERN).
The data concerns the anomalous electroweak production of
$WW$, $WZ$ and $ZZ$ boson pairs in association with two jets in
proton-proton collisions at $\sqrt{s}=13\ \TeV$ with a
luminosity of $35.9\pm0.9\ \mathrm{fb}^{-1}$ \cite{CMS:2019qfk}.
\begin{figure}[!ht]
	\centering
	\includegraphics[width=0.5\linewidth]{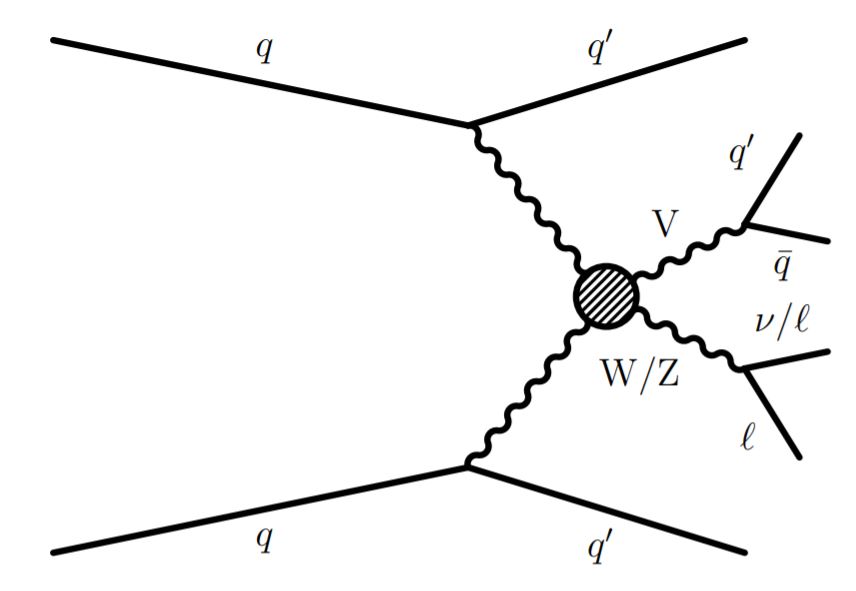}
	\caption{Feynman diagrams that involve triple and quadruple gauge couplings in proton-proton collisions \cite{CMS:2019qfk}.}
	\label{fig:diagrammi1}
\end{figure}
The diagrams contributing to this process are shown in
Fig.~\ref{fig:diagrammi1}. We return again to the convention in which
the Higgs doublet is represented as a vector with two complex
components (see Eq.~\eqref{eq:higgsdoublet}), and the Skyrme term as a
linear combination of the operators $Q_{\phi^4}^{(1)}$,
$Q_{\phi^4}^{(2)}$ and $Q_{\phi^4}^{(3)}$.
Their most general orthogonal linear combination is commonly denoted
as
\begin{equation}
    \label{eq:qgc}
	\begin{split}
		\mathcal{L}_{\phi^4}=&\ \frac{f_{S,0}}{M^4}Q_{\phi^4}^{(2)}+\frac{f_{S,1}}{M^4}Q_{\phi^4}^{(3)}+\frac{f_{S,2}}{M^4}Q_{\phi^4}^{(1)} \\
		=&\	\frac{\alpha}{M^4}\mathcal{O}_{Sk}+\frac{\beta}{M^4}\cdot \frac{1}{2}\left[Q_{\phi^4}^{(2)}-Q_{\phi^4}^{(1)}\right]+\frac{\gamma}{M^4}\cdot \frac{1}{2} \left[Q_{\phi^4}^{(1)}+Q_{\phi^4}^{(2)}+Q_{\phi^4}^{(3)}\right],
	\end{split}
\end{equation}
where $M$ is some mass scale. The three dimensionless coefficients
$\alpha$, $\beta$ and $\gamma$ are linear functions of $f_{S,0}$,
$f_{S,1}$ and $f_{S,2}$ 
\begin{equation}
  \alpha=\frac{f_{S,0}+f_{S,2}-2f_{S,1}}{3}\ , \qquad
  \beta=f_{S,0}-f_{S,2}\ , \qquad
  \gamma=\frac{2}{3}(f_{S,0}+f_{S,1}+f_{S,2})\ .
\end{equation}
By definition, the constant $\Lambda$ that multiplies
$\mathcal{O}_{Sk}$ in the SMEFT expansion is 
\begin{equation}
  \frac{1}{\Lambda^4}=\frac{1}{3}\left(\frac{f_{S,0}}{M^4}+\frac{f_{S,2}}{M^4}-\frac{2f_{S,1}}{M^4}\right)\ .
\end{equation}
The current upper bounds on $f_{S,0}$, $f_{S,1}$ and $f_{S,2}$ can be
used to find a corresponding lower bound on $\Lambda$. The data for
$f_{S,0}$ and $f_{S,1}$ from Ref.~\cite{CMS:2019qfk}, with $95\%$ CL,
are shown in the table below: 
\begin{center}
  \begin{tabular}{LLL}
    \hline 
    & WW\ (\TeV^{-4}) & WZ\ (\TeV^{-4}) \\
    \hline
    f_{S,0}/M^4 & [-2.7,2.7] & [-40,40] \\ 
    f_{S,1}/M^4 & [-3.3,3.4] & [-32,32] \\ 
    \hline 
  \end{tabular} 
\end{center}
However, to the best of our knowledge, there are no data concerning
$f_{S,2}$, nor any theoretical bound from e.g.~unitarity or positivity
that allow to express it as a function of $f_{S,0}$ and
$f_{S,1}$. This makes constraining $\Lambda$ an impossible task.  

In order to make sense of this data, we are forced to make some
additional assumptions that cannot be justified from first
principles. For example, we can assume that custodial symmetry is
satisfied also to this order in the SMEFT, which is equivalent to
imposing the constraint $f_{S,0}=f_{S,2}$. In this case 
\begin{equation}
\label{eq:Lambda}
	\frac{1}{\Lambda^4}=\frac{2}{3}\left(\frac{f_{S,0}}{M^4}-\frac{f_{S,1}}{M^4}\right)
        \quad\implies\quad \Lambda \gtrsim 0.71\ \TeV \ .
\end{equation}
Another possibility is to fix the ratios between the $f_{S,i}$ using a UV completion of the Skyrme term. In Ref.~\cite{Criado:2020zwu}, three different UV completions are given. The Skyrme term can be generated by an $SU(2)$ triplet vector boson $V^{\mu}_a$ with Lagrangian
\begin{equation}
  \mathcal{L}_{\rm UV}=\frac{1}{2}\left(D_{\mu}V_{\nu}^aD^{\nu}V^{a\mu}-D_{\mu}V_{\nu}^aD^{\mu}V^{a\nu}+M^2V_{\mu}^aV^{a\mu}\right)+g_VV_{\mu}^a2\ \mathrm{Im}\left(\phi^{\dagger}\sigma^aD^{\mu}\phi\right)\ .
\end{equation}
Computing the Wilson coefficients, one can show that
\begin{equation}
    f_{S,0}=f_{S,2}=-\frac{1}{2}f_{S,1} \quad\implies\quad \Lambda \gtrsim 0.74\ \TeV\ .
\end{equation}
Another possibility is to generate the Skyrme term through a scalar singlet $S$ with Lagrangian
\begin{equation}
    \mathcal{L}_{\rm UV}=-\frac{1}{2}S(D^2+M^2)S+\kappa_SS|\phi|^2\ .
\end{equation}
In this case
\begin{equation}
    f_{S,2}=f_{S,0}=0 \quad\implies\quad \Lambda\gtrsim 0.82\ \TeV\ .
\end{equation}
In general, due to the fact in Eq.~\eqref{eq:Lambda} there is the
fourth power of $\Lambda$, for any model that predicts ratios
$\alpha/f_{S,j}\sim \mathcal{O}(1)$ the corresponding bound for
$\Lambda$ is very close to $0.7-0.8\ \TeV$, except when
$f_{S,0}+f_{S,2}-2f_{S,1}$ is very close to zero. 

\subsection{Dark matter}
In this section we compute the relic abundance of cosmological
EWSs. We assume that: 
\begin{enumerate}
    \item The cosmological EWSs have been produced after the
      electroweak phase transition ($T_{\rm EW}\simeq 10^{15} K$). This
      assumption is necessary, as the classical solution does not
      exist in the unbroken phase of the theory. 
    \item The production was thermal i.e.~the EWSs, $S$, have
      been produced in reactions of the form $X+\bar{X}\to S+\bar{S}$
      where $X$ are Standard Model particles. These reactions, of
      course, conserve $B+L$. As we are neglecting those production
      channels that result in a net $B+L$ change, we are probably
      misestimating the abundance actually predicted by the model.
    \item The cosmological EWSs freeze out at $T\sim M/30$, and
      at that moment were non-relativistic. 
    \item The thermally averaged cross section of the EWSs can
      be roughly approximated by the geometric cross section
      $\expval{\sigma v}\sim\pi\lambda_{\text{Compton}}^2$ at
      $\Lambda\simeq 130$ (see Sec.~\ref{sec:qcorr}).
\end{enumerate}
Given these assumptions, we can compute the abundance with the
solution of the Boltzmann equation \cite{Kolb:1990vq, Baumann:2022mni}:
\begin{equation}
\label{eq:boltzmann}
  \Omega_Sh^2 \sim 0.1\left(\frac{x_f}{10}\right)\left(\frac{10}{g_*(M)}\right)^{1/2}\frac{10^{-8}\mathrm{GeV}^{-1}}{\expval{\sigma v}}\ ,
\end{equation}
where $x_f=M/T\sim 30$ at freeze out and $g_*(M)\sim 90$ is the
number of relativistic degrees of freedom at temperature $M$. The
results of this calculations is 
\begin{equation}
\label{eq:omegah}
  \Omega_Sh^2\sim 10^{-5}\ ,
\end{equation}
which is significantly smaller than the DM abundance
$\Omega_{DM}h^2\simeq 0.2$. It is instructive to compare our results
with those of Ref.~\cite{Criado:2020zwu}. Assuming the
$\Lambda$-dependence of the EWS mass and radius to be that in
Eq.~\eqref{eq:massradius} and applying the formula
\eqref{eq:boltzmann}, one obtains $\Omega_Sh^2\simeq
(\Lambda/3\ \mathrm{TeV})^4$, which, at the classical-to-quantum
transition $\Lambda\simeq 310\ \mathrm{GeV}$ is approximately
$\Omega_Sh^2\sim 10^{-4}$. We conclude that thermally produced EWSs
cannot be considered plausible DM candidates.

\section{Conclusion}
\label{sec:conclusion}

In this paper we discussed various aspects of the electroweak-Skyrme
theory. This model has been studied in the past, always with the
assumption of the Higgs field frozen at the vacuum expectation
value. Recently, this model has been reconsidered assuming a dynamical
Higgs field. Our finding is that the effect of the dynamical Higgs is
even greater than previously recognized. We find that in the
range of $\Lambda$ from $100$ to $300$ GeV, the EWS mass goes from
$350$ to $30$ GeV, roughly a factor 10 lower than what was
previously found. The profile functions for the EWS
show two regions: one internal where the Higgs field is vanishing and
the gauge fields are at their maximum value, and one external with no
gauge fields and where the Higgs approaches its vacuum expectation
value. As $\Lambda$ is increased the soliton becomes smaller and
lighter, as expected before.
We considered the rotational degrees of freedom and their
quantization. We found that in the region of parameters we are
considering, the gap in the rotational energies is of order of GeV,
thus considerably smaller than the EWS mass scale. The rotational
energy gap is big enough to justify the rigid rotor approximation we
have been using. We showed that, if no other terms are present in the
effective Lagrangian, the EWS must be quantized as a
boson and the ground state is a scalar particle with spin zero. For
the vibrational energy scales, we considered the breather (interpreted as a particle, it is often called an oscillon), a particular
vibration of the scale factor which is relatively easy to compute, and
should give the right order of magnitude of the vibrational scales.

We considered the effect of the hypercharge, as a perturbation of the
EWS. This gives a magnetic dipole moment to the
classical EWS. When the rotational degrees of freedom
are quantized, the magnetic dipole is oriented toward the spin
direction of the particle. In the spin-zero ground state, there is no
remnant of the magnetic dipole moment. We considered fermion
interactions with the electroweak Skyrmion. They can induce a fermion
number to the EWS through the Goldstone-Wilczek
mechanism. We considered the coupling of the EWS with
the vector bosons and the Higgs field, and then computed the large
distance interaction potential between two solitons.

We then consider, at the qualitative level, the issues of stability and
quantum corrections. The EWS is not a topologically
protected soliton, it is separated by the sphaleron barrier to the
normal vacuum. It is, at most, metastable. In the region of parameters
we considered, with $\Lambda$ above $100$ GeV. The mass of the
EWS is always well below the sphaleron barrier, so
for any phenomenologically reasonable value of $\Lambda$ we expect to
have a metastable soliton. Quantum corrections are expected to be very
important. At $\Lambda\simeq 200$ GeV the EWS is already at
the scale of the Higgs mass. In general, solitons can be treated
semi-classically only if they are heavier than any perturbative
particle. So we can trust a semiclassical soliton analysis in a
relatively small window, between the stability barrier
$\Lambda_{\rm min}$ (which we can estimate roughly to be 50 GeV) and the
region where quantum effects become considerable $\Lambda_{\rm max}$
(which we can estimate to be roughly 200 GeV). For larger values of 
$\Lambda$, the particle corresponding to the EWS is
certainly there, and it is metastable with a very large lifetime. We
cannot say much about its properties such as mass, radius and coupling
with other particles. A naive extrapolation of the semiclassical
analysis would predict that the mass goes to zero as $\Lambda \to \infty$,
but a conservative guess is that it will stabilize itself around the
Higgs mass scale.

Phenomenological constraints on $\Lambda$ are quite severe, and they
impose a parameter space that is outside the rigorous validity of the
approximations we have made. For $\Lambda$ in the range preferred by
phenomenology, the EWS is a very quantum soliton and also very
stable. This poses an interesting challenge for the future as this
particle should be there, at least for this choice of
higher-derivative corrections, and it is not semiclassical.
Another aspect to look into is the modification of the EWS sector as
we change the types of higher-derivative corrections.

Various things remain to be explored in the future.
One aspect that should be considered with more precision is
a complete analysis of the energy functional in the landscape of
field configurations, with the sphaleron and all the possible saddle
points, and how they are affected by the combination of Higgs mass
and Skyrme term.
To study the exact decay process, one should look, at least
semiclassically, at the bounce solution to understand precisely the
tunneling process that leads to the EWS decay.
The numerical approach, both for the landscape and for the bounce
problems, is quite complex to be carried out with the same precision
we obtained for the solution of the EWS and this would require
dedicated work in the future.
Another task for future work is to study multi-EWS bound states. The
multi-Skyrmions are very well studied in the Skyrme literature, and
are known to be very stable solutions with multiple baryon
numbers. They should also exist in the electroweak Skyrme model and
may have an impact on the phenomenological aspects.
From a phenomenological perspective, the most pressing thing to do now
would be to learn how to treat the EWS in a region of parameter space
where quantum corrections are large.

\subsubsection*{Acknowledgments}

We thank Paolo Panci for useful discussions, especially on the
phenomenological aspects of this problem.
We also thank Alfredo Urbano and  Antonio Junior Iovino for useful
discussions.
Finally, we thank the anonymous referee for questions on the
stability of the EWS.
The work of S.~B.~is supported by the INFN special project 
grant ``GAST''.
S.~B.~G.~thanks the Outstanding Talent Program of Henan University and
the Ministry of Education of Henan Province for partial support.
The work of S.~B.~G.~is supported by the National Natural Science
Foundation of China (Grants No.~11675223 and No.~12071111) and by the
Ministry of Science and Technology of China (Grant No.~G2022026021L).
The work of G.~S.~is supported by the INFN special project grant
``GAGRA''.

\appendix

\section{The breather}\label{app:breather}
The scaling law under dilatations of the EWS fields are expressed in
Eq.~\eqref{eq:scaling}, and the new time-component of the weak gauge field
is introduced in Eq.~\eqref{eq:timecomponent}. We plug these transformed
quantities into the Lagrangian and find an effective action for the
dilatation parameter $\eta$. The terms that do not depend on $W_0$ can
be easily found by a change of variables. Define
\begin{subequations}
    \begin{align}
        A&\equiv \int \d^3x\left(\frac{1}{4g^2}\mathrm{tr}\ W_{ij}^2-\frac{1}{8\Lambda^4}\mathrm{tr}(D_i\Phi^{\dagger}D_j\Phi-D_j\Phi^{\dagger}D_i\Phi)^2\right),\\
        B&\equiv \int \d^3x\ \frac{1}{2}\mathrm{tr}\ D_i\Phi^{\dagger}D_i\Phi,\\
       C&\equiv \int \d^3x\ \frac{\lambda}{4}(\mathrm{tr}\ \Phi^{\dagger}\Phi-v^2)^2\ ,
    \end{align}
\end{subequations}
then
\begin{equation}
  A\longmapsto \eta A\ , \qquad
  B\longmapsto \frac{B}{\eta}\ , \qquad
  C\longmapsto\frac{C}{\eta^3}\ .
\end{equation}
The virial law for the static solution reads
\begin{equation}
    A=B+3C\ .
\end{equation}
The remaining terms are the time-dependent terms:
\begin{subequations}
  \begin{align}
    \frac12D_1\frac{\dot\eta^2}{\eta^3} &\equiv \frac12\int d^3x\ W_{0i}^a W_{0i}^a
    -\frac{1}{8\Lambda^2}\int d^3x\ 2\tr(D_{[0}\Phi^\dag D_{i]}\Phi)^2\ ,\\
    \frac12D_2\frac{\dot\eta^2}{\eta^5} &\equiv \frac12\int d^3x\;\tr D_0\Phi^\dag D_0\Phi\ .
  \end{align}
\end{subequations}
We now make the rescaling
\begin{equation}
  r\to\frac{v}{\Lambda^2}r\ , \qquad
  t\to\frac{v}{\Lambda^2}t\ , \qquad
  (f_1,f_2,b,\xi)\to\frac{2}{g}(f_1,f_2,b,\xi)\ .
\end{equation}
We introduce the dimensionless Lagrangian for the breather
\begin{equation}
    L_{\text{br,nat}}=\frac{1}{2}\left(D_{1,\rm nat}+\frac{D_{2,\rm nat}}{\eta^2}\right)\frac{\dot{\eta}^2}{\eta^3}-\left(\eta A_{\text{nat}}+\frac{1}{\eta}B_{\text{nat}}+\frac{1}{\eta^3}C_{\text{nat}}\right)\ ,
\end{equation}
where the dimensionless coefficients obey
\begin{equation}
  (L_{\text{br}},A,B,C,D_1,D_2)=\frac{4\pi v^3}{\Lambda^2}(L_{\text{br,nat}},A_{\text{nat}},B_{\text{nat}},C_{\text{nat}},D_{1,\rm nat},D_{2,\rm nat})\ ,
\end{equation}
and are given by
\begin{subequations}
  \begin{align}
    A_{\text{nat}} &= \int_0^{\infty} \d r \bigg\{\alpha \left[\left(f_1'-2f_2\frac{b}{r}\right)^2+\left(f_2'+(2 f_1 - 1)\frac{b}{r}\right)^2+\frac{2}{r^2}\left(f_1^2+f_2^2 - f_1\right)^2\right] \non
    &\phantom{=\int_0^{\infty}\d r\bigg\{\ }
    +(f_1^2+f_2^2)\sigma^2\left[(\sigma')^2+\frac{\sigma^2}{r^2}\left(b^2+\frac{f_1^2+f_2^2}{2}\right)\right]\bigg\}\ ,\\
    B_{\text{nat}} &= \int_0^{\infty} \d r\left[\frac{r^2}{2}(\sigma')^2+\sigma^2\left(f_1^2+f_2^2+\frac{b^2}{2}\right)\right]\ ,\\
    C_{\text{nat}} &= \int_0^{\infty} \d r\beta r^2(\sigma^2-1)^2\ ,\\
    \label{eq:d1}
    D_{1,\rm nat} &= \int_0^{\infty} r^2 \d r\bigg\{2\alpha\left[\left(f_1' - \frac{2\xi f_2}{r}\right)^2+\left(f_2'-\xi\frac{1-2f_1}{r}\right)^2+\frac12(b'-\xi')^2\right] \non
    &\phantom{=\int_0^{\infty}r^2\d r\bigg\{\ }
    +2\sigma^2\left[(f_1^2+f_2^2)\left((\sigma')^2+\frac{\xi^2\sigma^2}{r^2}\right)+\frac{(\sigma')^2}{2}(\xi-b)^2\right]\bigg\}\ ,\\
    \label{eq:d2}
    D_{2,\rm nat} &= \int_0^{\infty} r^2 \d r\left(r^2(\sigma')^2 + \xi^2\sigma^2\right)\ .
\end{align}
\end{subequations}
$\alpha$ and $\beta$ are defined in Eq.~\eqref{eq:alphabeta}.
For small perturbations $\epsilon=\eta-1$ around the classical
minimum, the Lagrangian reduces to that of a harmonic oscillator with
frequency 
\begin{equation}
  \omega_{c}=\sqrt{\frac{2B_{\text{nat}}+12C_{\text{nat}}}{D_{1,\rm nat}+D_{2,\rm nat}}}\ .
\end{equation}
The new profile function $\xi$ can be determined as a function of
$f_1$, $f_2$, $b$ and $\sigma$. If we consider $\xi$ as a dynamical
variable of the system, it must satisfy the Euler-Lagrange equation 
\begin{equation}
  \left(\frac{\d}{\d r}\frac{\delta}{\delta \xi'}-\frac{\delta}{\delta \xi}\right)L_{\text{br,nat}}=0 \quad\implies\quad
  \left(\frac{\d}{\d r}\frac{\delta}{\delta \xi'}-\frac{\delta}{\delta \xi}\right)(D_{1,\rm nat}+D_{2,\rm nat})=0\ .
\end{equation}
Suitable boundary conditions are
\begin{equation}
    \xi(0)=0\ , \qquad \xi(\infty)=0\ .
\end{equation}
Explicitly, the equation of motion for $\xi$ is
\begin{equation}
    \begin{split}
      \xi'' - b''+\frac2r(\xi'-b')&=-\left(f_1' - \frac{2\xi f_2}{r}\right)\frac{4f_2}{r}+\left(f_2'-\xi\frac{1-2f_1}{r}\right)\frac{2(2f_1-1)}{r} \\
      &\phantom{=\ }
      +\frac{\sigma^2}{\alpha}\left[\xi+\frac{2\xi\sigma^2(f_1^2+f_2^2)}{r^2}+(\sigma')^2(\xi-b)\right]\ .
\end{split}
\end{equation}
An educated guess is that the solution of $\xi$ is $\xi=b$, since this
reduces the equation of motion to an algebraic equation that can
easily be compared to that for $b$.
We obtain
\begin{equation}
  -\left(f_1' - \frac{2b f_2}{r}\right)\frac{4f_2}{r}
  +\left(f_2' - b\frac{1-2f_1}{r}\right)\frac{2(2f_1-1)}{r}
  +\frac{\sigma^2}{\alpha}\left[
    b
    +\frac{2b\sigma^2(f_1^2+f_2^2)}{r^2}
    \right] = 0\ ,
\end{equation}
which can be solved for $b$:
\begin{equation}
b = \frac{2r\alpha(2f_2 f_1'+(1-2f_1)f_2')}{2\alpha\left((1-2f_1)^2+4f_2^2\right) + r^2\sigma^2 +2(f_1^2+f_2^2)\sigma^4}\ ,
\end{equation}
which is exactly the static equation of motion for $b$. This
establishes that $\xi=b$ is a solution. 
A quick inspection of Eqs.~\eqref{eq:d1}-\eqref{eq:d2} shows that
$\xi=b$ is a local minimum of $D_{1,\rm nat}+D_{2,\rm nat}$, and thus it maximizes
$\omega_{c}$.

\section{Interaction potentials}
\label{app:intpot}
In this appendix, we briefly show how to compute the interaction
potential between two solitons taking into account all the relevant
contributions. In order to avoid irrelevant complications, we consider
the case of a soliton made of a scalar field $\phi$ with action 
\begin{equation}
  S(\phi)=\int\d^4x\left[\frac{1}{2}\partial_{\mu}\phi\ \partial^{\mu}\phi-\frac{1}{2}m^2\phi^2+V(\phi)\right]\ .
\end{equation}
Let $\phi(\textbf{x})$ be a soliton field configuration. For
$|\textbf{x}|\gg R$, with $R$ being the soliton radius, the asymptotic
field $\phi_{\infty}(\textbf{x})$ satisfies the approximate equations
of motion
\begin{equation}
    (-\nabla^2+m^2)\phi_{\infty}(\textbf{x})=S(\textbf{x})\ ,
\end{equation}
where $S(\textbf{x})$ is a source with support in $\textbf{x}=0$. Let
$\phi(\textbf{x}-\textbf{x}_1)$ and $\phi(\textbf{x}-\textbf{x}_2)$ be
two identical solitons with $\textbf{x}_{12}\gg R$.
Far from $\textbf{x}_1$ and $\textbf{x}_2$, the total asymptotic field
$\phi_{\infty}(\textbf{x})=\phi_{\infty}(\textbf{x}-\textbf{x}_1)+\phi_{\infty}(\textbf{x}-\textbf{x}_2)$
satisfies
\begin{equation}
    (-\nabla^2+m^2)\phi_{\infty}(\textbf{x})=S(\textbf{x}-\textbf{x}_1)+S(\textbf{x}-\textbf{x}_2)\ .
\end{equation}
From these equations of motion, and neglecting anharmonic contributions, we can guess an approximate expression for the total energy of the two solitons
\begin{equation}
\label{eq:solenergy}
    E_{\mathrm{tot}}=2E_{\mathrm{sol}}+\int d^3\textbf{x}\left[\phi^{(1)}(-\nabla^2+m^2)\phi^{(2)}-\phi^{(2)}S^{(1)}-\phi^{(1)}S^{(2)}\right]+\ldots
\end{equation}
where we have used the notation
$\phi^{(i)}=\phi_{\infty}(\textbf{x}-\textbf{x}_i)$ and
$S^{(i)}=S(\textbf{x}-\textbf{x}_i)$ and where $E_{\mathrm{sol}}$ is
the energy of a single soliton. The ellipses contain anharmonic terms,
that can be neglected within our approximation. The first term of the
integrand in Eq.~\eqref{eq:solenergy} yields the \emph{tail-tail}
contributions, while the other two terms yield the \emph{core-tail}
contribution. Using the equations of motion, one ends up with 
\begin{equation}
  E_{\mathrm{tot}}=2E_{\mathrm{sol}}-\int d^3\textbf{x}\ \phi^{(1)}(-\nabla^2+m^2)\phi^{(2)}+\ldots
\end{equation}
which explains the minus sign in Eq.~\eqref{eq:interaction}. The
resulting interaction potential is always attractive, in agreement
with the particle exchange interpretation.

\section{Asymptotic behavior of the EWS fields}
\label{app:asymptotic_behavior}

\subsection{Fields at large distance}

The asymptotic behavior of the solutions can be easily found by
inserting the Ansatz \eqref{eq:asymp_ansatz} into the linearized equations of
motion. The results obtained with this method depend only on the
quadratic part of the action and on the form of the Ansatz 
\beq
\label{eq:eomlarge}
\left(-\nabla^2+m_h^2\right)\Phi(\textbf{x})=0\ , \qquad
\left(-\nabla^2+m_{W}^2\right)W_{i}(\textbf{x})=0 \ , \qquad
\partial_iW_i(\textbf{x})=0\ .
\eeq
The linearized equations \eqref{eq:eomlarge} must be valid for
$\textbf{x}\neq 0$.
As shown in App.~\ref{app:intpot} their right-hand side can 
contain a distribution with support at the origin. Solving the
linearized equations for the Higgs field, we simply obtain 
\begin{equation}
\label{eq:asymphiggs}
	\sigma(r)\underset{r\to \infty}\longrightarrow 1+\frac{c_h}{m_h}\frac{e^{-m_h r}}{4\pi r} \ , \qquad \sigma(r)\underset{r\to 0}\longrightarrow d_h\frac{e^{m_hr}-e^{-m_hr}}{m_hr} \ ,
\end{equation}
where $c_h$ and $d_h$ are real constants. For the weak gauge fields,
the linearized equations of motion for the functions
$F_1=\frac{f_1}{r}$, $F_2=\frac{f_2}{r}$ and $B=\frac{b}{r}$ are
coupled 
\begin{subequations}
	\begin{equation}
	\label{eq:asympcons}
		 \hat{x}_a\left[B'+\frac{2}{r}(B-F_2)\right]=0 \ ,
	\end{equation}
	\begin{equation}
	\label{eq:weaklinearized}
		\begin{split}
		 	\varepsilon_{aij}\hat{x}_j\left[F_1''+\frac{2}{r}F_1'-\frac{2}{r^2}F_1-m_{W}^2F_1\right]+\hat{x}_i\hat{x}_a\left[B''+\frac{2}{r}B'-\frac{4}{r^2}(B-F_2)-m_{W}^2B\right] \\
			\mathop+(\delta_{ia}-\hat{x}_i\hat{x}_a)\left[F_2''+\frac{2}{r}F_2'+\frac{2}{r^2}(B-F_2)-m_{W}^2F_2\right] =0 \ .
		\end{split}
	\end{equation}
\end{subequations}
The independence of the three members of the second equation can be
easily proven by multiplying them by $\hat{x}_i$,
$(\delta_{ik}-\hat{x}_i\hat{x}_k)$ and
$\varepsilon_{aik}\hat{x}_k$. The first equation is a constraint on
the initial conditions, so it introduces no redundancy in the
system. The most general solution of these equations at infinity is 
\begin{equation}
\label{eq:asymptotic}
    \begin{split}
	     F_1(r)&\underset{r\to\infty}\longrightarrow \frac{c_{\alpha}}{4\pi}\left[j_{-2}(\i m_Wr)+\i y_{-2}(\i m_Wr)\right] \ , \\
	    B(r)+2F_2(r) &\underset{r\to\infty}\longrightarrow \frac{c_{\beta}}{4\pi}\frac{e^{-m_Wr}}{m_Wr} \ , \\
	    B(r)-F_2(r)&\underset{r\to\infty} \longrightarrow -\i\frac{c_{\gamma}}{4\pi}\left[j_{-3}(\i m_Wr)+\i y_{-3}(\i m_Wr)\right] \ ,
    \end{split}
\end{equation}
where $c_{\alpha}$, $c_{\beta}$, $c_{\gamma}$ are real constants and
$j_{\ell}$ and $y_{\ell}$ are spherical Bessel functions. Instead, for
$r\rightarrow 0$, we have 
\begin{equation}
    \begin{split}
	    F_1(r)&\underset{r\to 0}\longrightarrow \i d_{\alpha}\, y_{-2}(\i m_Wr) \ , \\
	    B(r)+2F_2(r)& \underset{r\to 0}\longrightarrow d_{\beta}\frac{e^{m_Wr}-e^{-m_Wr}}{m_Wr}\ ,\\
	    B(r)-F_2(r)&\underset{r\to 0} \longrightarrow d_{\gamma}\, y_{-3}(\i m_Wr) \ ,
    \end{split}
\end{equation}
where $d_{\alpha}$, $d_{\beta}$, $d_{\gamma}$ are real constants. The
constraint \eqref{eq:asympcons} has not been imposed yet.
This will be done in the next subsection by passing to momentum space.

\subsection{Source terms}
The source term for the Higgs field $h=v(\sigma-1)$ can be easily
obtained from the distributional identity
$(-\nabla^2+m^2)\frac{e^{-mr}}{4\pi r}=\delta^{(3)}(\textbf{x})$
applied to the field in Eq.~\eqref{eq:asymphiggs}.
The result is 
\begin{equation}
    S^{(h)}(\textbf{x})=\frac{c_hv}{m_h}\delta^{(3)}(\textbf{x})\ .
\end{equation}
We now search for a source term for the weak fields $S^{(W)} $
satisfying 
\begin{equation}
    \label{eq:source}
    \left(-\nabla^2\delta_{ij}+\partial_i\partial_j+m_W^2\delta_{ij}\right)W_j(\textbf{x})=S^{(W)}_i(\textbf{x})\ .
\end{equation}
The source $S^{(W)}_i$ can be a function of the derivatives $\partial_i$
and of the Pauli matrices $\tau^a$ applied on a Dirac delta centered
at the origin, $\delta^{(3)}(\textbf{x})$, and must satisfy all the
symmetries of $W_i$. In order to find the source, we make the
Fourier transform of both sides of Eq.~\eqref{eq:source} using
the same notation of Eq.~\eqref{eq:boundary}. First, we rearrange
the terms in Eq.~\eqref{eq:profile} as 
\begin{equation}
    W_{ai}=\varepsilon_{aij}\hat{x}_jF_1+\left(\hat{x}_i\hat{x}_a-\frac{1}{3}\delta_{ia}\right)(B-F_2)+\frac{1}{3}\delta_{ia}(B+2F_2)\ .
\end{equation}
Then, we separate the radial and angular part of the
Fourier-transformed weak field by means of the identity 
\begin{equation}
    e^{-\i\textbf{p}\cdot\textbf{x}}=\sum_{\ell}(2\ell+1)(-\i)^{\ell}P_{\ell}(\hat{p}\cdot\hat{x})j_{\ell}(pr)\ ,
\end{equation}
where $P_{\ell}$ are the Legendre polynomials and $j_{\ell}$ are the modified spherical Bessel functions. For the angular part, we use the identities
\begin{subequations}
    \begin{equation}
        \int\d\Omega_{\textbf{x}}P_{\ell}(\hat{x}\cdot\hat{p})=4\pi\delta_{\ell,0}\ ,
    \end{equation}
    \begin{equation}
        \int\d\Omega_{\textbf{x}}\hat{x}_iP_{\ell}(\hat{x}\cdot\hat{p})=\frac{4\pi}{3}\delta_{\ell,1}\hat{p}_i\ ,
    \end{equation}
    \begin{equation}
        \int\d\Omega_{\textbf{x}}\left(\hat{x}_i\hat{x}_j-\frac{1}{3}\delta_{ij}\right)P_{\ell}(\hat{x}\cdot\hat{p})=\frac{4\pi}{5}\delta_{\ell,2}\left(\hat{p}_i\hat{p}_j-\frac{1}{3}\delta_{ij}\right)\ .
    \end{equation}
\end{subequations}
We find that
\begin{align}
    \widetilde{W}_{ai}(\textbf{p})=&-4\pi\i\varepsilon_{aij}\hat{p}_j\int_0^{\infty}r^2\d r\ F_1(r)j_1(pr)+\frac{4\pi}{3}\delta_{ia}\int_0^{\infty}r^2\d r\ (B(r)+2F_2(r))j_0(pr) \non
    &-4\pi\left(\hat{p}_i\hat{p}_j-\frac{1}{3}\delta_{ij}\right)\int_0^{\infty}r^2\d r\ (B(r)-F_2(r))j_2(pr)\ .
\end{align}
Then we substitute the asymptotic expressions for $F_1$, $F_2$ and $B$
into the equation and use the integrals 
\begin{subequations}
    \begin{equation}
        \int_0^{\infty}r^2\d r\ \frac{e^{-ar}}{ar}j_0(qr)=\frac{1}{a}\frac{1}{q^2+a^2}\ ,
    \end{equation}
    \begin{equation}
        \int_0^{\infty}r^2\d r\ \left(j_{-2}(\i ar)+\i y_{-2}(\i ar)\right)j_1(qr)=\frac{q}{a^2}\frac{1}{q^2+a^2}\ ,
    \end{equation}
    \begin{equation}
        \int_0^{\infty}r\d r\ \left(j_{-3}(\i ar)+\i y_{-3}(\i ar)\right)j_2(qr)=\frac{\i q^2}{a^3}\frac{1}{q^2+a^2}\ .
    \end{equation}
\end{subequations}
We finally obtain the result
\begin{equation}
\begin{split}
    \widetilde{W}_{ai}(\textbf{p})=-\frac{1}{p^2+m_W^2}\left[\i\frac{c_{\alpha}}{m_W^2}\varepsilon_{aij}p_j-\frac{1}{3}\delta_{ia}\frac{c_{\beta}}{m_W}+\left(p_ip_a-\frac{1}{3}\delta_{ia}p^2\right)\frac{c_{\gamma}}{m_W^3}\right]\ .
\end{split}
\end{equation}
Before applying the operator
$(p^2\delta_{ij}-p_ip_j+m_W^2\delta_{ij})$, we need to impose the
condition $\partial_iW_i^a(\textbf{x})=0$, which, in momentum space,
takes the form 
\begin{equation}
    0=p_i\widetilde{W}_{ai}(\textbf{p})=\frac{1}{p^2+m_W^2}\left[\frac{1}{3}\frac{c_{\beta}}{m_W}-\frac{2c_{\gamma}}{3m_W^3}p^2\right]p_a \quad\implies\quad c_{\beta}=c_{\gamma}=0\ ,
\end{equation}
which means that the source is
\begin{subequations}
\begin{align}
    \widetilde{S}_{ai}^{(W)}(\textbf{p})&=-\i\frac{c_{\alpha}}{m_W^2}\varepsilon_{aij}p_j\ , \qquad \text{(momentum space)}\\
    S_{ai}^{(W)}(\textbf{x})&=-\frac{c_{\alpha}}{m_W^2}\varepsilon_{aij}\partial_j\delta^{(3)}(\textbf{x})\ . \qquad \text{(position space)}
\end{align}
\end{subequations} 
The asymptotic expression for $W_{ai}$ in position space is thus
\begin{equation}
  W_{ai}(\textbf{x})=-\frac{c_{\alpha}}{m_W^2}\varepsilon_{aij}\partial_j\frac{e^{-m_Wr}}{4\pi r}\ .
\end{equation}

\phantomsection
\addcontentsline{toc}{section}{References}
\bibliography{references}
\bibliographystyle{JHEP}

\end{document}